\documentclass[journal,12pt,onecolumn,draftclsnofoot,]{IEEEtran}
\usepackage{amsmath,amsfonts}
\usepackage{algorithmic}
\usepackage{algorithm}
\usepackage{array}
\usepackage{textcomp}
\usepackage{stfloats}
\usepackage{url}
\usepackage{verbatim}
\usepackage{graphicx}
\usepackage{cite}
\usepackage{cite}
\usepackage{amsmath,amssymb,amsfonts}
\usepackage{algorithmic}
\usepackage{graphicx}
\usepackage{textcomp}
\usepackage{xcolor}
\usepackage{subcaption}
\usepackage{booktabs}
\usepackage{siunitx}
\def\BibTeX{{\rm B\kern-.05em{\sc i\kern-.025em b}\kern-.08em T\kern-.1667em\lower.7ex\hbox{E}\kern-.125emX}}
\usepackage[prependcaption]{todonotes}
\presetkeys%
{todonotes}%
{inline}{}
\usepackage[utf8]{inputenc}
\usepackage[T1]{fontenc}
\usepackage{grffile}
\graphicspath{{figures/}}
\usepackage{bbm}
\usepackage{url}
\usepackage{etoolbox}
\newtoggle{isit}
\newtoggle{arxivlink}
\newtoggle{trimmer}
\iftoggle{isit}{}{%
  \usepackage{hyperref}
}%
\usepackage[shortlabels]{enumitem}
\usepackage{hyperref}    
\usepackage[a-1b]{pdfx}  
\usepackage{algorithm}
\usepackage{algorithmic}  

\newcommand{\defeq}{\triangleq}
\newcommand{\reals}{\mathbb{R}}
\newcommand{\extReals}{\bar{\reals}}
\renewcommand{\Pr}{\operatorname{\mathbb P}}
\newcommand{\E}{\operatorname{\mathbb E}}

\newcommand{\ind}[2]{\left(#1\colon #2\right)} 

\newtheorem{exam}{Example}
\newtheorem{definition}{Definition}
\newtheorem{remark}{Remark}
\newtheorem{prop}{Proposition}
\newtheorem{cor}{Corollary}
\newtheorem{lemma}{Lemma}

\newtheorem{assump}{Assumption}

\newcommand{\indicator}[1]{I_{#1}}
\newcommand{\set}[2]{\left\{#1\colon #2\right\}}
\renewcommand{\d}{\mathop{}\!d}
\newcommand{\uiuc}{University of Illinois Urbana–Champaign}

\newcommand{\pmfLike}{a}
\newcommand{\pmfInfty}{b}

\newcommand{\ROC}{\mathsf{ROC}}
\newcommand{\AUC}{\mathsf{AUC}}
\renewcommand{\hat}{\widehat}
\renewcommand{\phi}{\varphi}
\newcommand{\levy}{L\'{e}vy}
\newcommand{\ROCE}{\hat\ROC_{\mathrm E}}
\newcommand{\ROCCE}{\hat\ROC_{\mathrm{CE}}}
\newcommand{\ROCML}{\hat\ROC_{\mathrm{ML}}}
\newcommand{\AUCML}{\hat\AUC_{\mathrm{ML}}}
\newcommand{\figWidth}{0.47}

\begin{document}

\title{Maximum Likelihood Estimation of Optimal Receiver Operating Characteristic Curves From Likelihood Ratio Observations\\

}

\author{%
  \IEEEauthorblockN{Bruce Hajek and Xiaohan Kang,}
  \IEEEauthorblockA{\uiuc\\
    Electrical and Computer Engineering and Coordinated Science Laboratory\\
    Urbana, Illinois\\
    Email: b-hajek@illinois.edu, xkang515@gmail.com} }
\maketitle

\begin{abstract}
  The optimal receiver operating characteristic (ROC) curve, giving the maximum probability of detection as a function of the probability of false alarm, is a key information-theoretic indicator of the difficulty of a binary hypothesis testing problem (BHT).  It is well known that the optimal ROC curve for a given BHT, corresponding to the likelihood ratio test, is determined by the probability distribution of the observed data under each of the two hypotheses.  In some cases, these two distributions may be unknown or computationally intractable, but independent samples of the likelihood ratio can be observed.  This raises the problem of estimating the optimal ROC for a BHT from such samples.  The maximum likelihood estimator of the optimal ROC curve is derived, and it is shown to converge almost surely to the true optimal ROC curve in the \levy\ metric, as the number of observations tends to infinity.  Finite sample size bounds are obtained for three other estimators: the classical empirical estimator, based on estimating the two types of error probabilities from two separate sets of samples, and two variations of the maximum likelihood estimator called the split estimator and fused estimator, respectively.  The maximum likelihood estimator is observed in simulation experiments to be considerably more accurate than the empirical estimator, especially when the number of samples obtained under one of the two hypotheses is small.  The area under the maximum likelihood estimator is derived; it is a consistent estimator of the area under the true optimal ROC curve.
\end{abstract}

\thanks{A portion of this work appeared in {\em Proceeding of the IEEE International Symposium on Information Theory,} 2022.}

\section{Introduction}

Consider a binary hypothesis testing problem (BHT) with observation $X$.  The observation $X$ could be high dimensional with continuous and/or discrete components.  Suppose $g_0$ and $g_1$ are the probability densities of $X$ with respect to some reference measure, under hypothesis $H_0$ or $H_1$, respectively.  Then the likelihood ratio is $R = \frac{g_1(X)}{g_0(X)}$.  By the Neyman–Pearson lemma, the optimal decision rule for a specified probability of false alarm, is to declare $H_1$ to be true if either $R > \tau$ or ($R = \tau$ and a biased coin comes up heads) for a suitable threshold $\tau$ and bias of the coin.  The optimal receiver operating characteristic (ROC) curve, giving the maximum probability of detection as a function of the probability of false alarm, is a key information-theoretic indicator of the difficulty of the BHT.  Because we focus on the optimal ROC, which is determined by the BHT rather than the specific decision rule, we use the terms ``optimal ROC'' and ``ROC'' interchangeably.

This paper addresses the problem of estimating the ROC curve for a BHT from independent samples $R_1, \dots , R_n$ of the likelihood ratio.  Specifically, we assume for some deterministic sequence, $\ind{I_i}{i \in [n]}$, that $R_i$ is generated from an instance of the BHT such that hypothesis $H_{I_i}$ is true.  This problem can arise if the densities $g_0$ and $g_1$ are unknown, but can be factored as $g_k(x)=u(x)h_k(x)$ for $k\in\{0,1\}$, for some unknown (or very difficult-to-compute) function $u$ and known functions $h_0$ and $h_1.$ Then the likelihood ratio can be computed for an observation $X$ using $R = \frac{h_1(X)}{h_0(X)},$ but the distribution of the likelihood ratio depends on the unknown function $u.$ So if it is possible, through simulation or repeated physical trials, to generate independent instances of the BHT, it may be possible to generate the independent samples $R_1, \dots , R_n$ as described.

To elaborate a bit more, we discuss a possible specific scenario related to Cox's notion of partial likelihood \cite{Cox75}.  Suppose $X = (Y_1,S_1,Y_2,S_2,\dots , Y_T,S_T),$ where the components themselves may be vectors.  The full likelihood under hypothesis $H_k$ for $k=0,1$ is the product of two factors given below, each of which is a product of $T$ factors: \iftoggle{trimmer}{\vspace{-0.2cm}}{}%
\begin{align*}
 \left( \prod_{t=1}^T  f_{Y_t|Y^{t-1}, S^{t-1}} (y_t|y^{t-1},s^{t-1};k) \right)
 \cdot \left( \prod_{t=1}^T  f_{S_t|Y^{t}, S^{t-1}} (s_t|y^{t},s^{t-1};k) \right)
\end{align*}
\iftoggle{trimmer}{\vspace{-0.3cm}}{}\\
where $y^t\defeq\ind{y_{t'}}{t'\in[t]}$.  Cox defined the first factor to be the partial likelihood based on $Y$ and the second factor to be the partial likelihood based on $S$.  If the first factor is very complicated but does not depend on $k$, and the second factor is known and tractable, we arrive at the form of the total likelihood described above: $g_k(x)=u(x)h_k(x)$ for
\iftoggle{trimmer}{$k = 0, 1$.}{$k\in\{0,1\}$.}   See \cite{KangHajek21} for a more detailed example application. 

To avoid possible confusion, we emphasize that the problem considered is an inference problem with independent observations, where the ROC is to be estimated.  The space of ROCs is infinite-dimensional.  We do not focus on finding the optimal decision rule for a BHT, which is already known to be the likelihood ratio test.

There is a large literature on ROC curves dating to the early 1940s.  Much of the emphasis relating to estimating ROC curves is focused on estimating the area under the ROC curve (AUC), a key performance measure for machine learning algorithms \cite{Bradley97}.  For estimation of the ROC curves, a popular approach is the binormal model such that the distribution of an observed score is assumed to be a monotonic transformation of a Gaussian random variable under either hypothesis, and maximum likelihood (ML) estimates of the parameters of the Gaussian distribution are found.  See \cite{MetzPan99,HsiehTurnbull96} and references therein.  The papers \cite{HsiehTurnbull96,Darlington73,Bamber75} and others address estimation of ROC curves from samples of ``scores'' or ``diagnostic variables'' that are assumed to have different distributions under the two hypothesis.  However, there is no assumed relationship between the two distributions; the distributions are not necessarily distributions of likelihood ratios. We have not found previous work on estimating ROC curves from likelihood ratio observations.

The first estimator we consider for the ROC curve, which we call the ``empirical ROC curve,'' is described by that name in \cite{DeLongDeLongClarke-Pearson88}, although that paper refers to ``diagnostic variables.'' The empirical ROC curve is the same up to a rotation as the ``sample ordinal dominance graph'' defined in \cite{Darlington73} and used in \cite[p.~400]{Bamber75}.   The bound and its proof that we show are close to those in \cite{HsiehTurnbull96}.  We view this as a known baseline estimator, and the contribution of our paper is to provide an alternative, if not better, estimator, by exploiting the strong relationship between the distributions of the likelihood ratio samples under the two hypotheses. Our use of \levy\ metric and the concavified empirical estimator may be new.

The next estimator we consider is the maximum likelihood estimator, which is the choice of ROC curve that maximizes the likelihood of the observed likelihood ratios.  There is an extensive literature on the maximum likelihood estimation method, dating back over one hundred years to R.A. Fisher \cite{Aldrich97}.  In the context of this paper the parameter to be estimated is infinite dimensional -- an ROC curve -- so that the theory of maximum likelihood estimation is largely not applicable.  Thus there is no {\em a priori} reason for the ML estimator of the ROC to have some strong properties.  But often ML estimators have nice properties and it is worth including them in the search for good estimators.  For example, the empirical estimator of a CDF based on samples generated from the CDF is the maximum likelihood estimator of the CDF.  And for estimation of ROCs based on likelihood ratio samples, we find that the ML estimator has an interesting form, is consistent, and performs rather well in simulations.

Consistency of an estimator means that as the number of observations converges to infinity for a fixed parameter, the estimator converges to the parameter in a suitable sense (in probability or almost surely, for example).  Consistency is widely considered to be an important property of an estimator because it implies accuracy with high probability as the number of samples converges to infinity \cite{Cramer46}.  Consistency of an estimator does not give bounds on accuracy for a finite number of observations.  Thus, it is important to find finite sample performance guarantees for estimators which can be used, for example, to make confidence intervals.  While we have not been able to produce satisfactory finite sample performance guarantees for the maximum likelihood estimator, we have found such bounds for variations of the estimator we call the split and fused estimators.

To our knowledge, the following are new contributions of this paper.  The formulation and identification of the maximum likelihood (ML) ROC estimator based on likelihood ratio observations, the proof of consistency of the ML estimator, a mapping ${\cal M}$ used in our proof of consistency, the formulation of two estimators closely related to ML, and the proof of finite sample size performance guarantees for those other two estimators.  In addition, we provide simulation results suggesting that the ML estimator and its variations are more accurate than the empirical estimators.

The paper is organized as follows.  Some preliminaries about ROC curves are given in Section~\ref{sec:formulation}.  The empirical estimator of the optimal ROC curve, based on using the empirical estimators for the two types of error probabilities, is considered in Section~\ref{sec:empirical_estimator}.  A performance guarantee is derived based on a well-known bound for empirical estimators of CDFs.  The ML estimator of the ROC curve is given in Section~\ref{sec:mle} together with a proof of its consistency.  A key tool is a mapping $\cal M$ from the set of all distributions supported on $[0,\infty]$ to the set of ROC curves.  The area under the ML estimator of the ROC curve is derived and is shown to be a consistent estimator of AUC.  In Section~\ref{sec:split_estimator}, two variations of the ML estimator, called the split estimator and fused estimator, are derived, and finite sample size performance bounds are given for them.  Simulations comparing the accuracy of the empirical and ML estimators are given in Section~\ref{sec:simulations}, and conclusions and future directions are in Section~\ref{sec:conclusions}.  Proofs are found in the appendix.

\section{Preliminaries About Optimal ROC Curves}
\label{sec:formulation}
\subsection{An extension of a cumulative distribution function (CDF)}
\label{sec:notation}
The CDF $F$ for an extended random variable $R$ (i.e., $R$ can take the value $\infty$) is defined by $F(\tau)=\Pr\{R\leq \tau\}$ for $\tau\in\reals$. The corresponding complementary CDF is defined by $F^c(\tau)=1-F(\tau)=\Pr\{R>\tau\}.$ In this paper $\infty$ always means $+\infty$.  Given a CDF $F$ with $F(0-) = 0$ and possibly a point mass at $\infty$, we define an extended version of $F$, and abuse notation by using $F$ to denote both $F$ and its extension.  The extension is defined for $\tau\in \reals\cup\{\infty\}$ and $\eta\in[0, 1]$, by $F(\tau, \eta) = (1- \eta) F(\tau-) + \eta F(\tau)$, where $F(\infty-) = \lim_{\tau\to\infty}F(\tau)$ and $F(\infty) = 1$.  Let $F(\{\tau\}) = F(\tau) - F(\tau-)$ denote the mass at $\tau$.  Thus, if $R$ is an extended random variable with CDF $F$, then $F(\tau,\eta)=\Pr\{R<\tau\} + \eta\Pr\{R = \tau\}$.  Note the extended version of $F$ is continuous and nondecreasing in $(\tau, \eta)$ in the lexicographic order with $F(0, 0) = 0$ and $F(\infty, 1) = 1$, and hence surjective onto $[0, 1]$.  Also, let the extended complementary CDF for $F$ be defined by $F^c(\tau, \eta) = 1- F(\tau,\eta)$, so that $F^c(\tau, \eta) = \Pr\{ R>\tau \} +(1- \eta) \Pr\{ R=\tau\}$.

\subsection{The optimal ROC curve for a BHT}
\label{sec:model}
Consider a BHT and let $F_0$ denote the CDF of the likelihood ratio $R$ under hypothesis $H_0$ and let $F_1$ denote the CDF of the observation $R$ under hypothesis $H_1$.  Then $\d F_1(r) = r\d F_0(r)$ for $r\in (0,\infty)$%
\iftoggle{isit}{}{ (see Appendix~\ref{app:relation} for details) },
and $F_1(0)=F_0(\{\infty\})=0$, while it is possible that $F_0(0)>0$ and/or $F_1(\{\infty\})>0$.

The likelihood ratio test with threshold $\tau$ and randomization parameter $\eta$ declares $H_0$ to be true if $R < \tau$, declares $H_1$ to be true if $R > \tau$, and declares $H_1$ to be true with probability $\eta$ if $R = \tau$.  The {\em optimal ROC curve} is the graph of the function $\ROC(p): 0\leq p \leq 1$ defined by $\ROC(p)=F^c_1(\tau, \eta)$ where $\tau$ and $\eta$ are selected such that $F^c_0(\tau, \eta) = p$.  This is well-defined because $F_0$ is surjective and for any $\tau$, $\tau'$, $\eta$, and $\eta'$ we have $F_0^c(\tau, \eta) = F_0^c(\tau', \eta')$ if and only if $F_1^c(\tau, \eta) = F_1^c(\tau', \eta')$.  Equivalently, the optimal ROC curve is the set of points traced out by $P = (F^c_0(\tau,\eta), F^c_1(\tau,\eta))$ as $\tau$ and $\eta$ vary.

\begin{prop}
  \label{prop:equivalence of Fs and ROC}
  Any one of the functions $F_0$, $F_1$, or $\ROC$ determines the other two.
\end{prop}
\begin{remark}
\begin{enumerate}
\item  $\ROC$ is a continuous, concave, nondecreasing function over $[0, 1]$ with $\ROC(0)\ge 0$ and $\ROC(1) = 1$.  Conversely, any such function is an ROC curve of some BHT.
\item In view of Proposition~\ref{prop:equivalence of Fs and ROC}, the BHT with likelihood ratio observations can be specified by fixing any one of the three components $F_0,F_1$ or $\ROC.$ We keep that in mind but use the triplet $(F_0,F_1,\ROC)$ to denote a BHT.  Since we deal exclusively with likelihood ratio observations we leave the phrase ``likelihood ratio'' out of the notation.
\end{enumerate}
\end{remark}

\subsection{The \levy\ metric}
Let ${\cal L}$ denote the set of nondecreasing functions mapping $\reals\to \reals\cup\{-\infty\}$ such that for each $A\in {\cal L}$ there are finite constants $c_0$ and $c_1$ such that $A(x)=-\infty$ for $x<c_0$ and $A(x)=A(c_1)>-\infty$ for $x\geq c_1.$ The {\em \levy\ distance} between $A,B \in {\cal L}$ is the infimum of $\epsilon > 0$ such that
\begin{align*}
  A(p-\epsilon) - \epsilon \leq B(p) \leq A(p+\epsilon)+\epsilon\quad\text{for all } p\in\reals,
\end{align*}
with the convention $-\infty \leq -\infty.$ A geometric interpretation of $L(A, B)$ is that it is the smallest value of $\epsilon$ such that the graph of $B$ is contained in the region bounded by the following two curves: An upper curve obtained by shifting the graph of $A$ to the left by $\epsilon$ and up by $\epsilon$, and a lower curve obtained by shifting the graph of $A$ to the right by $\epsilon$ and down by $\epsilon$.  If $A$ is a nondecreasing function defined over $[0,1]$ we extend it to a function in ${\cal L}$ by setting $A(x)=-\infty$ for $x<0$ and $A(x)=A(1)$ for $x\geq 1.$ For two such functions $A$ and $B$, $L(A,B)$ is defined to be the \levy\ distance of their extensions in $\cal L.$
\begin{remark}
  For nondecreasing functions on the interval $[0,1]$ it is easy to see the \levy\ metric is dominated by the $L_\infty$ metric $L_\infty(A, B) \defeq \sup_{p\in[0, 1]}|A(p) - B(p)|$.  Note that the \levy\ metric is $1/\sqrt{2}$ times the $L_\infty$ metric on $A$ and $B$ after rotating the graphs clockwise by $45$ degrees, and hence tolerates horizontal deviation better than $L_\infty$.  To see this, consider the ideal ROC curve $\ROC \equiv 1$ over $[0,1]$ and an estimate $\hat{\ROC}(p) = \min\{cp, 1\}$ for $p\in [0,1],$ where $c>0.$ Then for large $c$ the $L_\infty$ distance between them is $1$, while the \levy\ distance $\frac 1{c + 1}$ is small.
\end{remark}

\begin{lemma}
  \label{lemma:BHT_metric_bound}
  Let $F_{a,0}, F_{a,1}, F_{b,0}, F_{b,1}$ denote CDFs for probability distributions on $[0, \infty]$.  Let $A$ be the function defined on $[0,1]$ determined by $F_{a,0}, F_{a,1}$ as follows.  For any $p\in[0, 1]$, $A(p) = F_{a, 1}^c(\tau, \eta)$, where $(\tau, \eta)$ is the lexicographically smallest point in $[0,\infty]\times[0,1]$ such that $F_{a, 0}^c(\tau, \eta) = p.$ (If $F_{a,0}$ and $F_{a,1}$ are the CDFs of the likelihood ratio of a BHT, then $A$ is the corresponding optimal ROC.) Let $B$ be defined similarly in terms of $F_{b,0}$ and $F_{b,1}.$ Then
  \iftoggle{trimmer}{%
      \begin{equation}
        L(A, B) \leq \sup_{\tau \in [0,\infty), i\in\{0, 1\}} |F_{a,i}(\tau)-F_{b,i}(\tau)|.
        \label{eq:bound_on_L_dist}
      \end{equation}
    }{%
      \begin{align}
        L(A,B)
        \leq \sup_{\tau \in [0,\infty)} \max\{ |F_{a,0}(\tau)-F_{b,0}(\tau)|, |F_{a,1}(\tau)-F_{b,1}(\tau)| \}.
          \label{eq:bound_on_L_dist}
      \end{align}
    }%
\end{lemma}


\section{The Empirical Estimator of the ROC}
\label{sec:empirical_estimator}

\begin{figure}[htb]
  \centering
  \includegraphics[width=\figWidth\textwidth]{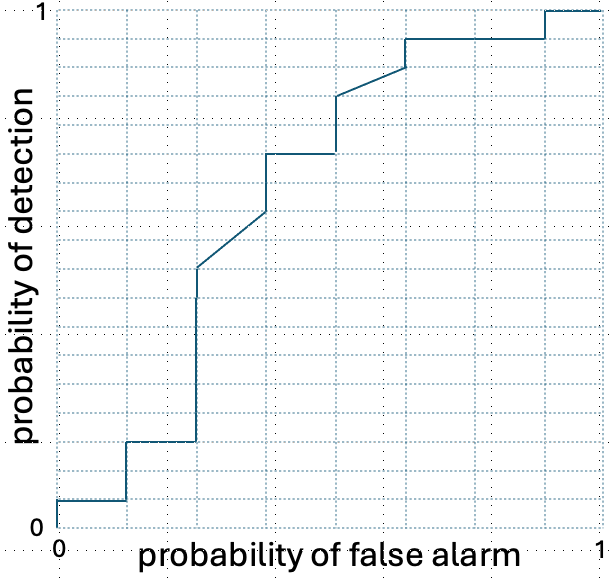}
  \caption{The ROC for the empirical estimator with 8 likelihood ratio samples drawn under $H_0$ and 18 under $H_1.$  Reading from the upper right corner of the figure indicates that the types of the rank-ordered samples are 01001\{01\}11011\{011\}11111101101 (i.e., the first, third, and fourth smallest samples are from $H_0.$  There is a tie between two samples, one from each hypothesis, for the sixth smallest sample.  And so on.).}
  \label{fig:ROC_empirical}
\end{figure}

Fix a BHT $(F_0,F_1,\ROC)$ and suppose for some positive integers $n_0$ and $n_1$ that independent random variables $R_{0,1}, \dots, R_{0,n_0},$ $ R_{1,1}, \dots , R_{1,n_1}$ are observed such that $R_{k,i}$ has CDF $F_k$ for $k=0,1$ and $1\leq i \leq n_k.$ A straight forward approach to estimate $\ROC$ is to estimate $F_k$ using only the $n_k$ observations having CDF $F_k$ for $k=0,1.$ In other words, let
\begin{align*}
  \hat{F_k} (\tau)= \frac 1 {n_k} \sum_{i=1}^{n_k} \indicator{\{R_{k,i}\leq \tau\}}
\end{align*}
for $k=0,1$ and let $\ROCE$, the {\em empirical estimator} of $\ROC$, have the graph swept out by the point $(\hat{F_0}^c(\tau,\eta), \hat{F_1}^c(\tau,\eta))$ as $\tau$ varies over $[0, \infty]$ and $\eta$ varies over $[0, 1]$.  In general, $\ROCE$ is a step function with all jump locations at multiples of $\frac 1 {n_0}$ and the jump sizes being multiples of $\frac 1 {n_1}$.  Moreover, $\ROCE$ depends on the numerical values of the observations only through the ranks (i.e., the order, with ties accounted for) of the observations, as illustrated in Fig.~\ref{fig:ROC_empirical}.

The estimator $\ROCE$ as we have defined it is typically not concave, and is hence typically not the optimal ROC curve for a BHT.  This suggests the {\em concavified empirical estimator} $\ROCCE$, defined to be the least concave majorant of $\ROCE$.  Equivalently, the region under the graph of $\ROCCE$ is the convex hull of the region under $\ROCE$.

We write ``$X_n \to c$ a.s.\ as $n\to\infty$'' where a.s.\ is the abbreviation for ``almost surely,'' to mean $\Pr\{\lim_{n\to\infty} X_n = c\} = 1$.  The following proposition provides some performance guarantees for the empirical and concavified empirical estimators.  The proof is based on the Dvoretzky–Kiefer–Wolfowitz (DKW) inequality with the optimal constant proved by Massart, which states that for any positive constant $\delta$, positive integer $n$, and CDF $F$, if $\hat{F}$ denotes the empirical CDF of $n$ independent samples from $F$, then
\begin{align}
  \label{eq:DKWM}
  \Pr\{ d_{KS}(F,\hat{F}) \geq \delta \} \leq 2e^{-2n\delta^2},
\end{align}
where $d_{KS}(F,G)$ denotes the Kolmogorov–Smirnov (KS) distance between CDFs $F$ and $G$:
\begin{align*}
  d_{KS} \stackrel{\triangle}{=} \sup_{c\in\reals} |F(c)-G(c)|.
\end{align*}
\begin{prop}
  \label{prop:dkw-empirical}
  Let $n=n_0+n_1$ and $\alpha = \frac{n_1}{n_1+n_0}$.  For any $\delta>0$ the empirical estimator satisfies
  \begin{align}
    \label{eq:empirical_estimator_bnd}
    \Pr\{ L(\ROC,\ROCE) \geq \delta \} \leq 2e^{-2n\alpha\delta^2} + 2e^{-2n(1-\alpha)\delta^2}.
  \end{align}
  Moreover, if $\alpha \in (0, 1)$ is fixed and $n_k\to\infty$ for $k=0, 1$ with $\frac{n_1}{n_0} = \frac{\alpha} {1-\alpha}$, then $L(\ROC, \ROCE)\to 0$ a.s.\ as $n\to\infty$.  In other words, $\ROCE$ is consistent in the \levy\ metric.  In general, $L(\ROC,\ROCCE) \leq L(\ROC,\ROCE)$, so the above statements are also true with $\ROCE$ replaced by $\ROCCE$.
\end{prop}

\begin{remark}
  A consistency result for the empirical estimator in terms of the uniform norm with some restrictions on the distributions $F_0$ and $F_1$ has been developed in \cite{HsiehTurnbull96}, similarly using the DKW inequality.  In particular, there is a bounded slope assumption not needed here because we use the \levy\ distance.
\end{remark}

While the bound \eqref{eq:empirical_estimator_bnd} seems reasonably tight for $\alpha$ near $1/2,$ the bound is degenerate if $\alpha$ is very close to zero or one.  The maximum likelihood estimator derived in the next section is consistent even if all the observations are generated under a single hypothesis, and the related split estimator has a finite sample performance guarantee stronger than the above for the empirical estimator if $|\alpha - \frac 1 2| > 0.12.$

\section{The ML Estimator of the ROC}
\label{sec:mle}

\subsection{Description of the ML ROC estimator. }
\label{sec:MLROC_description}

Consider a BHT and let $F_k$ denote the CDF of the likelihood ratio $R$ under hypothesis $H_k$ for $k=0,1,$ and suppose for some $n\geq 1$ and deterministic binary sequence $I_i: i\in [n]$, independent random variables $R_1, \dots, R_n$ are observed such that for each $i\in [n]$, the distribution of $R_i$ is $F_{I_i}.$ The likelihood of the set of observations is determined by $F_0$ and $F_1$, and hence, by Proposition~\ref{lemma:BHT_metric_bound}, also by $\ROC$ or by $F_0$ alone or by $F_1$ alone.  Hence, it makes sense to ask what is the maximum likelihood (ML) estimator of $\ROC$, or equivalently, what is the ML estimator of the triplet $(F_0,F_1,\ROC),$ given $I_i: i\in [n]$ and $R_i, i\in [n].$ The answer is given by Proposition~\ref{prop:ML_estimator} below.

Let $\phi_n$ be defined by
\begin{align}
  \label{eq:phi_n_def}
  \phi_n(\lambda)\defeq \frac 1 n \sum_{1\leq i \leq n: R_i <\infty} \frac 1 {1 - \lambda + \lambda R_i}.
\end{align}
Note that $\phi_n$ is finite over $[0,1)$, and continuous and convex over $[0,1]$.  Moreover, $\phi_n(1)= \infty$ if and only if $R_i =0$ for some $i$.

\begin{prop}
  \label{prop:ML_estimator}
  The ML estimator $(\hat{F}_{0,ML}, \hat{F}_{1,ML}, \ROCML)$ (or $(\hat{F}_0, \hat{F}_1, \ROCML)$ for short) is unique and is determined as follows.  $\ROCML$ is the optimal ROC curve corresponding to $\hat{F_0}$ and/or $\hat{F_1}$, where:
  \begin{enumerate}
  \item If $\frac 1 n \sum_{i = 1}^n R_i \leq 1$ (implying $R_i < \infty$ for all $i$), then for $\tau \in [0, \infty)$
    \begin{align*}
      \hat{F_0}(\tau) = \frac 1 n \sum_{i=1}^n \indicator{\{R_i\leq \tau\}};\quad\hat{F_1}(\tau) = \frac 1 n \sum_{i=1}^n \indicator{\{R_i\leq \tau\}} R_i.
    \end{align*}
  \item If $\frac 1 n \sum_{i = 1}^n \frac 1 {R_i } \leq 1$ (implying $R_i > 0$ for all $i$), then for $\tau \in [0, \infty)$
    \begin{align*}
      \hat{F_0}^c(\tau) = \frac 1 n \sum_{i=1}^n \indicator{\{R_i > \tau\}}\frac 1 {R_i };\quad\hat{F_1}(\tau) = \frac 1 n \sum_{i=1}^n \indicator{\{R_i\leq \tau\}}.
    \end{align*}
  \item If neither of the previous two cases holds, then for
    $\tau \in [0, \infty)$
    \begin{align*}
      \hat{F_0}(\tau) = \frac 1 n \sum_{i=1}^n \indicator{\{R_i\leq \tau\}} \frac 1 {1 - \lambda_n + \lambda_n R_i}%
      \iftoggle{trimmer}{,}{}%
    \end{align*}
    \iftoggle{trimmer}{}{%
        and }%
    \begin{align*}
      \hat{F_1}(\tau) = \frac 1 n \sum_{i=1}^n \indicator{\{R_i\leq \tau\}} \frac {R_i} {1 - \lambda_n + \lambda_n R_i},
    \end{align*}
    where $\lambda_n$ is the unique value in $(0,1)$ such that
    $\phi_n(\lambda_n) = 1$.
  \end{enumerate}
\end{prop}

\begin{remark}
  \label{remark:on_ML_estimator}
  \begin{enumerate}
  \item The estimator does not depend on the indicator variables $I_i: i\in [n].$  That is, the estimator does not take into account which observations are generated using which hypothesis.  For elaboration on this point, see Remark~\ref{remark:sufficient_statistic} below.
  \item Cases 1) and 2) can both hold only if $R_i = 1$ for all $i$, because $r + \frac 1 r \geq 2$ for $r\in [0, \infty]$ with equality if and only if $r = 1$.
  \item If case 1) holds with strict inequality, then $\hat{F_1}(\{\infty\}) > 0$, even though $R_i < \infty$ for all $i$.
  \item Similarly, if case 2) holds with strict inequality, then $\hat{F_0}(0) > 0$ even though $R_i >0$ for all $i$.
  \item Suppose case 3) holds.  The existence and uniqueness of $\lambda_n$ can be seen as follows.  Since case 2) does not hold, $\phi_n(1) > 1$.  If $R_i = \infty$ for some $i$ then $\phi_n(0) < 1$; and if $R_i < \infty$ for all $i$, then $\phi'_n(0) =\frac 1 n \sum_{i = 1}^n (1-R_i) < 0$, where we have used the fact case 1) does not hold.  Thus, in either case, $\phi_n(\lambda)<1$ if $\lambda > 0$ and $\lambda$ is sufficiently close to 0.  So $\phi_n$ is a convex function with an upcrossing of 1 in the interval $0 <\lambda < 1$, implying the existence and uniqueness of $\lambda_n$ in case 3.

  \item \iftoggle{isit}{}{%
        The proof of Proposition~\ref{prop:ML_estimator} is in Appendix~\ref{app:derivation_ML}.  }%
      Maximizing the likelihood is reduced to a convex optimization problem and the KKT conditions are used.
  \end{enumerate}
\end{remark}

The following corollary presents an alternative version of Proposition~\ref{prop:ML_estimator} that consolidates the three cases of Proposition~\ref{prop:ML_estimator}.  It is used in the proof of consistency of the ML estimator.

\begin{cor}
  \label{cor:ML_estimator_alt}
  The ML estimator is unique and is determined as follows.  For $\tau \in [0, \infty),$
  \begin{align*}
    \hat{F_0}^c(\tau) = \frac 1 n \sum_{i=1}^n \indicator{\{R_i > \tau\}} \frac 1 {1-\lambda_n + \lambda_n R_i}%
    \iftoggle{trimmer}{,}{}%
  \end{align*}
  \iftoggle{trimmer}{}{and}%
  \begin{align*}
    \hat{F_1}(\tau) = \frac 1 n \sum_{i=1}^n \indicator{\{R_i\leq \tau\}} \frac {R_i} {1 - \lambda_n + \lambda_n R_i},
  \end{align*}
  where $\lambda_n = \max\{ \lambda \in [0,1] : \phi_n(\lambda) \leq 1\}$.
\end{cor}

\begin{remark}
  By Corollary~\ref{cor:consistency} below, $\lambda_n\to\alpha$ a.s. if $F_0$ is not identical to $F_1$ and $\alpha$ is the fraction of samples with distribution $F_1.$ Thus, for $n$ large, $\lambda_n$ is approximately the prior probability $\alpha$ that a given observation is generated under hypothesis $H_1$ and $n\lambda_n$ is approximately the number of observations generated under $H_1$.  The ML estimator $\hat{F}_0$ can be written as
  \begin{align*}
    \hat{F_0}^c(\tau) = \frac 1 { n(1- \lambda_n)} \sum_{i=1}^n \indicator{\{R_i > \tau\}} \frac {1-\lambda_n} {1-\lambda_n + \lambda_n R_i},
  \end{align*}
  where $\frac {1-\lambda_n} {1 - \lambda_n + \lambda_n R_i}$ can be interpreted as an estimate of the posterior probability that $R_i$ was generated under $H_0$.
\end{remark}

\begin{remark}  \label{remark:sufficient_statistic}
  The factorization used in the proof of Proposition~\ref{prop:ML_estimator} suggests that, in general, the sample labels are not very useful in the context of estimating the ROC.  Another consequence of the factorization can be given as follows.  For clarity in this remark, we restrict attention to the case that both distributions have densities supported on $(0,\infty)$, but the idea works in general.  To apply the theory of sufficient statistics, we assume that $I = (I_1, \dots, I_n)$ is random with some known probability mass function $p_I.$ The parameter to be estimated is $\theta = \ROC,$ which determines the densities $f_0$ and $f_1$ with $f_1(r)=rf_0(r).$ With $R=(R_1, \dots, R_n)$, the observation is $(I,R).$ The density of the observations given $\theta$ can be written as the product of two factors: $f(I,R;\theta) =\left( p_I(I) \prod_i R_i^{I_i} \right) \left( \prod_i f_0(R_i) \right).$ The first factor is a function of the observation $(I,R)$ and does not include $\theta$, and the second factor is a function of $\theta$ and $R$.  Therefore, by the Fisher--Neyman factorization theorem, $R$ is a sufficient statistic for estimation of $\ROC$ given data $(I,R).$ The Rao--Blackwell theorem then implies that for any loss function that is convex in $\widehat{ROC},$ for the purpose of minimizing the expected loss, one can restrict attention to estimators $\widehat{\ROC}$ that only depend on $R$ and the distribution $p_I.$ For example, with $L$ denoting \levy\ distance, the loss functions $\widehat{\ROC}\mapsto L(\ROC, \widehat{\ROC})$ and $\widehat{\ROC}\mapsto e^{\eta L(\ROC, \widehat{\ROC})}$ for $\eta > 0$ are convex, where it is understood that linear combinations of ROC curves are taken after rotating clockwise 45 degrees (i.e., averaging along lines of slope $-1$).  So to minimize $E[L(\ROC, \widehat{\ROC})]$ or $E[e^{\eta L(\ROC, \widehat{\ROC})}]$ for $\eta > 0,$ over all estimators $\widehat{\ROC},$ one can restrict attention to estimators that depend only on $R$ and the assumed distribution $p_I.$ The expected loss $E[e^{\eta L(\ROC, \widehat{\ROC})}]$ is closely related to the DKW bound and performance guarantees in Section~\ref{sec:split_estimator}.
\end{remark}

\subsection{The mapping $\cal M$ and consistency of the ML estimator}
\label{sec:consistency}

The ML estimator is a mapping from the empirical CDF of the likelihood ratio observations to a BHT.  We shall prove consistency of the ML estimator by extending the domain of the mapping to the set of all CDFs of probability distributions supported on $[0,\infty]$ and showing that the resulting mapping $\cal M$ is continuous.   This also gives a way to interpret the ML estimator.

Given a BHT= $(F_0,F_1,\ROC)$ and a value $\alpha \in [0,1]$ let $F=(1-\alpha)F_0 + \alpha F_1.$ Then $F$ is the CDF of the likelihood ratio for an observation that is generated using $F_0$ with probability $1-\alpha$ and distribution $F_1$ with probability $\alpha.$ For $0< r < \infty$ it follows that $dF(r)=(1-\alpha) dF_0(r) + \alpha dF_1(r)$, which together with $dF_1(r)=rdF_0(r)$, gives rise to the following expressions for $F_0$ and $F_1$ in terms of $\alpha$ and $F.$
\begin{align}
  F_0^c(\tau) = \int_{\tau+}^\infty \frac 1 {1- \alpha + \alpha r} dF(r)
  \label{eq:alphaF0_fromF}\\
  F_1(\tau) = \int_0 ^\tau \frac r {1- \alpha + \alpha r} dF(r)
  \label{eq:alphaF1_fromF}
\end{align}

The following defines the mapping $\cal M$ from a set of CDFs to the set of BHT problems.
\begin{definition}
  Given a CDF $F$ for a probability distribution supported on $[0,\infty],$ let ${\cal M}(F) = (F_0,F_1,\ROC)$, where $(F_0,F_1, \ROC)$ is the BHT problem specified as follows.  Let
  \begin{align}
    \label{eq:phi_def}
    \phi(\lambda) = \int_0^\infty \frac 1 {1-\lambda + \lambda r}  dF(r)
  \end{align}
  and $\beta=\max\{\lambda\in [0,1]: \phi(\lambda) \leq 1\}.$  Then for $\tau \in [0,\infty),$ let
  \begin{align}
    F_0^c(\tau) = \int_{\tau+}^\infty \frac 1 {1- \beta + \beta r} dF(r)
       \label{eq:F0_fromF}\\
    F_1(\tau) = \int_0 ^\tau \frac r {1- \beta + \beta r} dF(r)
         \label{eq:F1_fromF}
  \end{align}
  Finally, let $\ROC$ denote the optimal ROC for the BHT determined by $F_0$ or, equivalently, by $F_1.$
\end{definition}

The following proposition proved in the appendix shows that any probability distribution on $[0,\infty]$ is the probability distribution of the likelihood function for some uniquely determined BHT and some prior probabilities $(1-\alpha, \alpha)$ on the hypotheses.

\begin{prop}
  \label{prop:M_properties}
  (i) Given a BHT $(F_0,F_1,\ROC),$ a value $\alpha \in [0,1],$ and $n\geq 1$, let $F=(1-\alpha)F_0 + \alpha F_1$ and suppose observations $R_1, \dots, R_n$ are independent with distribution $F$ and empirical distribution $\hat{F}.$ Then ${\cal M}(\hat{F}) = (\hat{F}_{0,ML}, \hat{F}_{1,ML}, \ROCML )$ and ${\cal M}(F)= (F_0,F_1,\ROC).$
  (ii) The mapping ${\cal M}:F\mapsto (F_0,F_1,\ROC)$ is continuous, using the Kolmogorov–Smirnov metric for $F, F_0$ and $F_1$ and the \levy\ metric for $\ROC.$ In addition, the variable $\beta$ associated with $\cal M$ is also continuous in $F$ over the set of all CDFs excluding the CDF $F(\{1\})=1.$
\end{prop}
\begin{remark}
A key challenge in proving part (ii) of Proposition~\ref{prop:M_properties} (in the appendix) is to show that if $d_{KS}(F,F_n)\to 0$ then $\phi_n \to \phi$ and $\beta_n\to \beta,$ where $\phi_n$ and $\beta_n$ arise in the definition of ${\cal M}(F_n)$ just as $\phi$ and $\beta$ arise in the definition of ${\cal M}(F).$
\end{remark}

We explain next how Proposition~\ref{prop:M_properties} implies consistency of the ML estimator.  Given a BHT=$(F_0,F_1,\ROC),$ and a value $\alpha \in [0,1]$ let $F=(1-\alpha)F_0 + \alpha F_1.$ Suppose the observations $R_1, R_2, \dots$ are independent, identically distributed random variables with CDF $F.$ Proposition~\ref{prop:M_properties} shows that the true BHT is equal to ${\cal M}(F).$ The DKW bound implies that $d_{KS}(\hat{F},F)\to 0$ a.s.  so by continuity of $\cal M$ the ML estimator ${\cal M}(\hat{F})$ converges to the true BHT, given by ${\cal M}(F).$ This implies the following corollary to Proposition~\ref{prop:M_properties}.

\begin{cor}[Consistency of $\ROCML$]
  \label{cor:consistency}
  $L(\ROCML, \ROC) \to 0$ a.s.\ as $n\to\infty$.  In addition, $d_{KS}(\hat{F}_{k,ML}, F_k)\to 0 ~~a.s.$ for $k\in\{0,1\}$ and, if $F(\{1\})\neq 1$, then $\lambda_n \to \alpha$~~a.s.
\end{cor}

\begin{remark}  \label{rem:non-Lip}
  Although Proposition~\ref{prop:M_properties} shows that the mapping $\cal M$ is continuous, Example~\ref{exam:not_Lip} in the Appendix shows that $\cal M$ is not Lipschitz continuous.  Therefore, straightforward application of the DKW inequality \eqref{eq:DKWM} does not pass through $\cal M$ in a simple way.  In theory $\cal M$ provides the following confidence bound:
  \begin{align*}
    \Pr\{ \hat{\ROC}_{ML} \in B_{\delta} \} \geq 1 - 2e^{2n\delta^2}
    ~~~~\mbox{where}~~B_{\delta} = \left\{{\cal M}(G):d_{KS}(G,\hat{F}) \leq \delta   \right\},
  \end{align*}
  and by the continuity of $\cal M$ it holds that $B_{\delta}$ shrinks down to $\ROCML$ as $\delta\to 0.$ It would be interesting to compute $B_{\delta}$ or find a tractable outer bound for it.
\end{remark}

\subsection{Area Under the ML ROC Curve}
\label{sec:AUROC_ML}

The area under $\ROCML$,
\iftoggle{trimmer}{denoted}{which we denote}
by $\AUCML$, is a natural candidate for an estimator of $\AUC$, the area under $\ROC$%
\iftoggle{trimmer}{.}{ for the BHT.}
An expression for it is given in the following proposition.  Let $\lambda_n$ be defined as in Corollary~\ref{cor:ML_estimator_alt} and for $i, i'\in [n]$, let
\begin{align*}
  T_{i,i'} = \frac{ \max\{R_i,R_{i'} \} } {2(1 - \lambda_n + \lambda_n R_i)(1 - \lambda_n + \lambda_n R_{i'})},
\end{align*}
with the following understanding.  Recall that if $R_i=0$ for some $i\in [n]$ then $\lambda_n < 1$, so the denominator in $T_{i,i'}$ is always strictly positive.  Also recall that if $R_i=\infty$ for some $i\in [n]$ then $\lambda_n > 0$, and the following is based on continuity: If $R_i=R_{i'}=\infty$ set $T_{i,i'}=0$.  If $R_i < R_{i'}=\infty$, set $T_{i,i'} = \frac 1 {2(1 - \lambda_n + \lambda_nR_i)\lambda_n}$.

\begin{prop}
  \label{prop:AUC_formulas}
  \begin{enumerate}
  \item The area under $\ROCML$ is given by
    \begin{align}
      \label{eq:AUChat_formula}
      \AUCML = \frac 1 {n^2}  \sum_{i=1}^n \sum_{i'=1}^n T_{i,i'}.
    \end{align}
  \item The estimator $\AUCML$ is consistent: $\AUCML \to \AUC$ a.s.\ as $n\to\infty$.
  \item Let $R, R'$ be independent random variables and use $\E_0$ to denote expectation when they both have CDF $F_0$.  Then
    \begin{align}
      \AUC& = \frac 1 2 \E_0[\max\{R,R'\}] + F_1(\{\infty\})  \label{eq:AUC_formula_1} \\
          & =  1 - \frac 1 2 \E_0[\min\{R,R'\}].    \label{eq:AUC_formula_2}
    \end{align}
  \item For $i\neq i'$, $\E[T_{i,i'}^{(\alpha)}]= \AUC$, where $T_{i,i'}^{(\alpha)}$ is the same as $T_{i,i'}$ with $\lambda_n$ replaced by $\alpha$.
  \end{enumerate}
\end{prop}

\begin{remark}
  \begin{enumerate}
  \item The expression \eqref{eq:AUChat_formula} can be verified by checking that it reduces to \eqref{eq:AUC_formula_1} in case $\E_0$ is replaced by expectation using $\hat{F_0}$ and $F_1$ is replaced by $\hat{F_1}$.  A more direct proof of \eqref{eq:AUChat_formula} is given.
  \item The true $\AUC$ for the BHT is invariant under swapping the two hypotheses.  Similarly, $\AUCML$ is invariant under replacing $\lambda_n$ by $1-\lambda_n$ and $R_i$ by $\frac 1 {R_i}$ for all $i$.  If $R_i=1$ for all $i$, $\AUCML = 1/2$.
  \item Part 4) of the proposition is to be expected due to the consistency of $\AUCML$ and the law of large numbers, because if $n$ is large, most of the $n^2$ terms in \eqref{eq:AUChat_formula} are indexed by $i, i'$ with $i\neq i'$, and we know, if $F_0$ is not identical to $F_1$, that $\lambda_n \to \alpha$ a.s.\ as $n\to\infty$.
  \end{enumerate}
\end{remark}

\section{The split and fused estimators of the ROC}
\label{sec:split_estimator}

As noted in Remark~\ref{rem:non-Lip} above, since the mapping $\cal M$ is not Lipschitz continuous, the method of directly using the DKW inequality does not work to give a good finite sample bound for the ML ROC estimator.  The difficulty is related to pinning down the value of $\lambda_n$ satisfying $\lambda_n = \max\{ \lambda \in [0,1] : \phi_n(\lambda) \leq 1\}$ for the function $\phi_n$ depending on the data.  In order to obtain estimators with a finite sample size performance bound, we relax our requirement somewhat and assume the estimator can depend on a parameter $\lambda$ which, for the performance evaluation, is assumed to equal the parameter $\alpha$, equal to the prior probability that any given sample is from $H_1.$

Given samples $R_1 \leq \dots \leq R_n$ the ML estimator $\ROCML$ can be described as follows.  It is constructed by placing end-to-end $n$ line segments such that the $i^{th}$ segment has slope $R_i$, horizontal displacement $\frac 1 {n (1 - \lambda_n+\lambda_n R_i)},$ and vertical displacement $\frac {R_i} {n (1 - \lambda_n+\lambda_n R_i)}.$ The segments are adjoined from left to right in the order of nonincreasing slope.  If $0 < \lambda_n < 1$ then the sums of the horizontal and vertical displacements are both one so the ROC can be anchored at each end by the points (0,0) and (1,1).

If the value $\lambda_n$ is replaced by some other value $\lambda$ then it is not possible to anchor such graph at both (0,0) and (1,1).  So instead, we consider two functions that we call pseudo ROCs, the first obtained by anchoring the function on the upper right at (1,1) and the second obtained by anchoring the function on the lower left at (0,0).

Specifically, given samples $R_1 \leq \dots \leq R_n$ and $\lambda \in [0,1]$ we define two pseudo ROC curves.  We assume that if $\lambda=0$ (corresponds to $H_0$ being true) then $R_i<\infty$ for all $i$ and if $\lambda=1$ (corresponds to $H_1$ being true) then $R_i>0$ for all $i.$ Under this assumption the horizontal and vertical displacements are well defined and finite.

Define ${\cal R}_{UR}(\hat{F},\lambda)$ to be the piecewise affine function over $\reals$ as follows, where $j_{\infty} = |\{i: R_i=\infty\}|:$
\begin{align}
  \label{eq:fUR_def}
  &{\cal R}_{UR}(\hat{F},\lambda)(p) \nonumber \\
  &= \begin{cases} -\infty & \text{if } p < 1- \frac 1 n \sum_{i=1}^n \frac{1}{1-\lambda + \lambda R_i},\\
    1 -\frac 1 n \sum_{i=1}^k \frac{R_i}{1-\lambda + \lambda R_i} & \text{if } p=1 -\frac 1 n \sum_{i=1}^k \frac{1}{1-\lambda + \lambda R_i} \mbox{ for some } 1 \leq k \leq n-j_{\infty}\\
    1 & \text{if } p\geq 1,
  \end{cases}
\end{align}
and ${\cal R}_{UR}(\hat{F},\lambda)$ is affine over the maximal intervals not covered by the righthand side of \eqref{eq:fUR_def}.  Similarly, for $\lambda \in [0,1]$ define ${\cal R}_{LL}(\hat{F},\lambda)$ to be the piecewise affine function as follows:
\begin{align}
  &{\cal R}_{LL}(\hat{F},\lambda)(p) \nonumber\\
  &= \begin{cases} -\infty & \text{if } p < 0,\\
    \frac{j_{\infty}}{n\lambda} & \text{if } p = 0\\
    \frac{j_{\infty}}{n\lambda} + \frac 1 n \sum_{i=k}^{n-j_{\infty}} \frac{R_i}{1-\lambda + \lambda R_i} & \text{if } p=\frac 1 n \sum_{i=k}^{n-j_{\infty} } \frac{1}{1-\lambda + \lambda R_i} \mbox{ for } 1 \leq k \leq n-j_{\infty} \\
    \frac{j_{\infty}}{n\lambda} + \frac 1 n \sum_{i=1}^{n-j_{\infty}} \frac{R_i}{1-\lambda + \lambda R_i} & \text{if } p\geq \frac 1 n \sum_{i=1}^{n-j_{\infty}} \frac{1}{1-\lambda + \lambda R_i} \\
  \end{cases}
\end{align}
The subscript ``UR'' reflects the fact that when restricted to the interval $(-\infty,1],$ the function ${\cal R}_{UR}(\hat{F},\lambda)$ is anchored at the upper right in the sense that ${\cal R}_{UR}(\hat{F},\lambda)(1)=1.$ Similarly, the subscript ``LL'' reflects that fact that when restricted to the interval $[0,\infty],$ the function ${\cal R}_{LL}(\hat{F},\lambda)$ is anchored at the lower left at $(0,\frac{j_{\infty}}{n\lambda})$ which is $(0,0)$ plus a vertical jump.  Note that ${\cal R}_{UR}(\hat{F},\lambda)$ and ${\cal R}_{LL}(\hat{F},\lambda)$ are translations of each other as graphs in $\reals^2$.  Both functions are concave functions in $\cal L.$

For a given $\hat{F}$ and $\lambda$, the function ${\cal R}_{UR}(\hat{F},\lambda)$ can fail to be a valid ROC curve because it is possibly negative in a subinterval of $[0,1].$ Similarly, ${\cal R}_{LL}(\hat{F},\lambda)$ can exceed one in an interval of $[0,1]$ or have value less than one at $p=1.$ We therefore define {\em clean} modifications of these two estimators so that the outputs are valid ROC curves, as follows.

Define $r^{min}(p)=p$ and $r^{max}(p)=1$ for $0\leq p \leq 1.$ Any (optimal) ROC curve must satisfy $r^{min} \leq \ROC \leq r^{max}$ over $[0,1]$ and must be concave. Let $T^{proj}$ be the operator that maps a function on $[0,1]$ to a function on $[0,1]$ with graph between those of $r^{min}$ and $r^{max}$:
\begin{align}
  T^{proj}f = \min\{ \max\{ f , r^{min}\}, r^{max} \},
\end{align}
and let $T^{conc}f$ denote the operator that maps a function $f$ on $[0,1]$ to the least concave majorant of $f$ over the interval $[0,1].$ The clean modifications are defined as ${\cal R}_{URC}(\hat{F},\lambda) = T^{conc}\circ T^{proj} \left( {\cal R}_{UR}(\hat{F},\lambda)\right)$ and ${\cal R}_{LLC}(\hat{F},\lambda) = T^{conc}\circ T^{proj} \left( {\cal R}_{LL}(\hat{F},\lambda)\right).$ These modifications are easily computed -- see Algorithm~\ref{alg:CUR} for the computation of ${\cal R}_{URC}(\hat{F},\lambda).$ The computation of ${\cal R}_{LLC}(\hat{F},\lambda)$ is the same up to symmetry.

\begin{algorithm}
  \caption{Algorithm to produce ${\cal R}_{URC}(\hat{F},\lambda)$ }
  \label{alg:CUR}
  \begin{algorithmic}
    \REQUIRE{$\lambda, n$, ordered likelihood ratio samples $R_1\leq \dots \leq R_n$}
    \STATE $p_0\gets 1~~~~~q_0 \gets 1~~~~~i \gets 0$
    \WHILE{()}
    \IF {$R_{i+1}>q_i/p_i$}
    \STATE $p_{i+1} \gets 0~~~~~q_{i+1} \gets 0~~~~~K\gets i+1$
    \STATE {\bf break} \COMMENT{escape while loop}
    \ENDIF
    \STATE $p_{i+1} \gets p_i - \frac 1 {n(1-\lambda)+ \lambda R_{i+1}} ~~~~~ q_{i+1} \gets q_i - \frac {R_{i+1}} {n(1-\lambda)+ \lambda R_{i+1}}$
    \IF {$p_{i+1}\leq 0$}
    \STATE    $p_{i+1} \gets 0~~~~~q_{i+1} \gets q_i - R_{i+1}*p_i~~~~~K\gets i+1$
    \STATE {\bf break}  \COMMENT{escape while loop}
    \ENDIF
    \ENDWHILE
    \RETURN $K$, representation points $(p_i,q_i)_{0\leq i \leq K}$ of ROC curve ${\cal R}_{URC}(\hat{F},\lambda)$
  \end{algorithmic}
\end{algorithm}

Define the {\em split ROC estimator} by
\begin{align*}
  {\cal R}_S(\hat{F},\lambda) = \left\{
  \begin{array}{cc}
    {\cal R}_{URC}(\hat{F},\lambda) & \mbox{if } 0 \leq \lambda \leq \frac 1 2\\
    {\cal R}_{LLC}(\hat{F},\lambda) & \mbox{if } \frac 1 2 \leq \lambda \leq  1.\\
  \end{array} \right.
\end{align*}

Define the {\em fused ROC estimator} ${\cal R}_F(\hat{F},\lambda)$ to be obtained by first rotating the graphs of ${\cal R}_{LLC}$ and ${\cal R}_{URC}$ clockwise by $45^o$, taking a convex combination of them, and then rotating counterclockwise by $45^o.$ More formally, ${\cal R}_F(\hat{F},\lambda)$ is defined to be the output of Algorithm~\ref{alg:fuse} for input $\left({\cal R}_{LLC}(\hat{F},\lambda),{\cal R}_{URC}(\hat{F},\lambda),\lambda\right).$ The ROC curves in the algorithm are piecewise linear and continuous, so each such function can be represented by a finite list of points on the graph of the function that include all the inflection points.  A rotation of the graph of such function can be represented by a rotation of the points in the finite list representing the graph.  The convex combination of two graphs can be accomplished by first adding breakpoints to either graph as necessary so the lists of points representing the two graphs have the same breakpoints.  The operation of rotating before taking the convex combination in the definition of ${\cal R}_F(\hat{F},\lambda)$ makes the definition symmetric between the two hypotheses and also allows us to obtain a tighter performance guarantee.
\begin{algorithm}
  \caption{Fusion of two ROC curves}
  \label{alg:fuse}
  \begin{algorithmic}
    \REQUIRE $\ROC_1,\ROC_2,\lambda\in [0,1]$
    \FOR{$k\in \{0,1\}$}
    \STATE $\widetilde{\ROC_k} \gets \mbox{Rotate}(\ROC_k,45^\circ\mbox{ clockwise})$
    \ENDFOR
    \STATE $\widetilde{\ROC} \gets\lambda \widetilde{\ROC}_1+ (1-\lambda) \widetilde{\ROC}_2$
    \STATE $\ROC \gets \mbox{Rotate}(\widetilde{\ROC},45^\circ\mbox{ counterclockwise})$
    \RETURN $\ROC$
  \end{algorithmic}
\end{algorithm}

The following proposition provides finite sample size performance guarantees for the four estimators of this section.
\begin{prop}
  \label{prop:finite_sample_bnds}
  Given a BHT triplet $(F_0,F_1,\ROC)$ and $\alpha\in [0,1],$ suppose $\hat{F}$ is the empirical CDF of samples $R_1, \dots , R_n$ independently generated using CDF $F=(1-\alpha)F_0 + \alpha F_1.$ Then
  \begin{align}
    \Pr\left\{L(\ROC, {\cal R}_{URC}(\hat{F},\alpha) ) \geq \delta\right\} & \leq 2\exp\left( - 2 n(1-\alpha)^2 \delta^2 \right)\label{eq:UR_est_bnd}\\
    \Pr\left\{L(\ROC, {\cal R}_{LLC}(\hat{F},\alpha) ) \geq \delta\right\} & \leq 2\exp\left( - 2 n\alpha^2 \delta^2 \right)\label{eq:LL_est_bnd} \\
    \Pr\left\{L(\ROC, {\cal R}_S(\hat{F},\alpha) ) \geq \delta\right\} & \leq 2\exp\left( - 2 n[\max\{\alpha, 1-\alpha\} \delta]^2 \right)\label{eq:S_est_bnd}\\
    \Pr\left\{L(\ROC, {\cal R}_{F}(\hat{F},\alpha) ) \geq \delta\right\} &\leq 2\exp(-n\delta^2/2)\label{eq:F_est_bnd}
  \end{align}
\end{prop}
\begin{remark}
The split estimator reduces to either the URC or LLC estimator, whichever one gives the better bound, so we won't discuss the URC and LLC estimators further.
The righthand sides of \eqref{eq:UR_est_bnd} - \eqref{eq:F_est_bnd} have the form $\exp(-nc\delta^2)$ where $c$ is a function of $\alpha.$   The bound \eqref{eq:empirical_estimator_bnd}  for the empirical estimator has two terms with the larger one also having the form $\exp(-nc\delta^2)$ for $\alpha = 2\min\{\alpha, 1- \alpha\}.$  We do not have a finite sample size upper bound for the ML estimator. The constants $c$ for the empirical, split, fused, and ML estimators are shown in Table~\ref{table:c_vs_alpha} and Figure~\ref{fig:c_vs_alpha}.
\begin{table}[h!]
  \caption{Constants $c$ vs. $\alpha$ in finite sample upper bounds.}
  \label{table:c_vs_alpha}
  \centering
  \begin{tabular}{SS} \toprule
    {estimator} & {c} \\  \midrule
    {empirical} & {$2\min\{\alpha, 1-\alpha\}$} \\
    {split} & {$2(\max\{\alpha, 1-\alpha \})^2$} \\
    {fused} & {0.5} \\
    {ML} & {} {n.a.} \\   \bottomrule
  \end{tabular}
\end{table}
\begin{figure}
  \centering
  \includegraphics[width=0.7\linewidth]{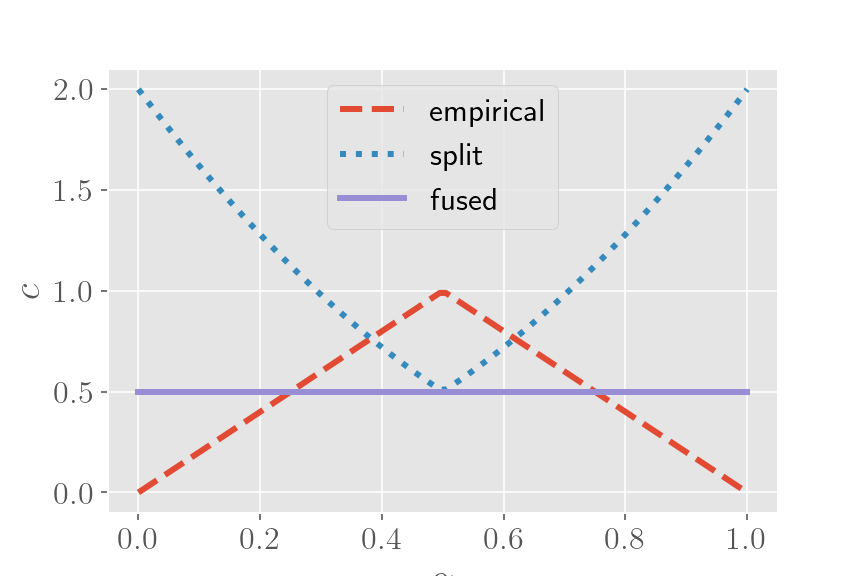}
  \caption{Constants $c$ vs.\ $\alpha$ in finite sample upper bounds.}
  \label{fig:c_vs_alpha}
\end{figure}
The bound \eqref{eq:S_est_bnd} for the split estimator is tighter than the bound \eqref{eq:empirical_estimator_bnd} for the empirical estimator if $\min\{\alpha, 1-\alpha\} < [\max\{\alpha, 1-\alpha\}]^2$ which holds if $0< \alpha < 0.38$ or $0.62<\alpha < 1.$ The split and fused estimators require use of $\alpha$ and use of the numerical values of the samples, but unlike the empirical estimator, they do not depend on which samples were generated under which hypothesis.

  If knowledge of $\alpha$ is not available, one idea is to first produce the estimate $\lambda_n$ associated with ${\cal M}(\hat{F})$ (since $\lambda_n \to \alpha$ a.s.) and plug $\lambda_n$ in for $\lambda$ in the split estimator or fused estimator.  But in either case, the resulting estimator would just be $\ROCML.$
\end{remark}

\section{Simulations}
\label{sec:simulations}
In this section we test the estimators in a simple binormal setting.  Let $X$ have the $\mathcal N(0, 1)$ distribution under $H_0$ and the $\mathcal N(\mu, 1)$ distribution under $H_1$.  Then the likelihood ratio for an observation $X$ is $R = \exp\left(\mu X - \frac 12\mu^2\right)$ and the ROC curve is given by $\ROC(p) = 1 - \Phi\left(\Phi^{-1}(1 - p) - \mu\right)$, where $\Phi$ is the CDF of the standard Gaussian distribution.  We first present the average \levy\ distance of the estimators from the true ROC and then present the distribution of the \levy\ distance of the estimators from the true ROC.

\begin{figure}[thbp]
  \begin{minipage}[b]{1.0\linewidth}
    \centering
    \includegraphics[width=\figWidth\textwidth]{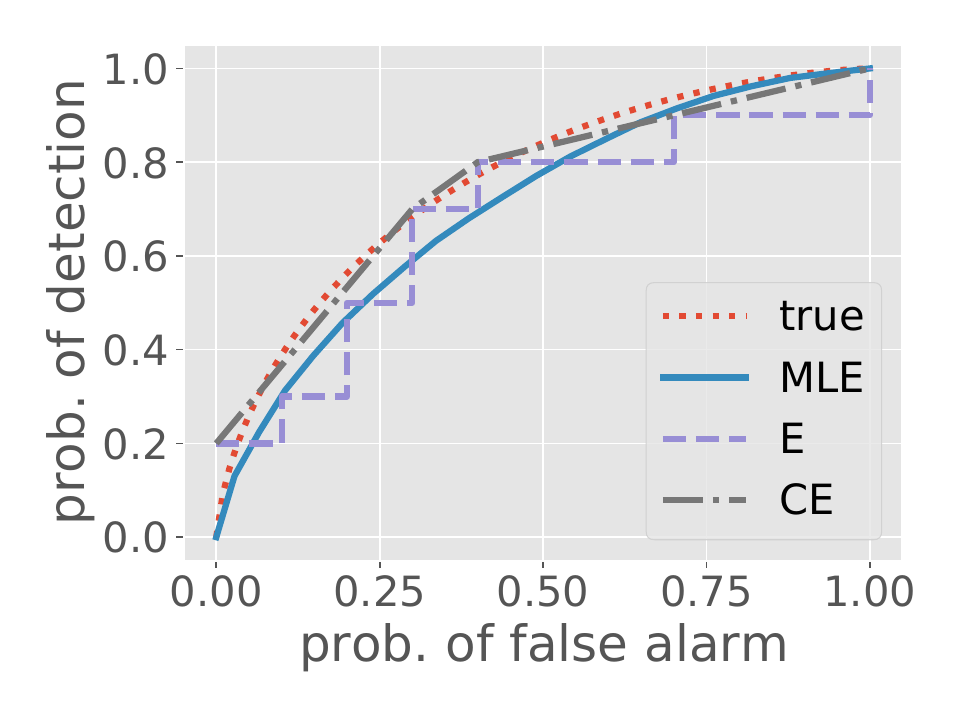}\includegraphics[width=\figWidth\textwidth]{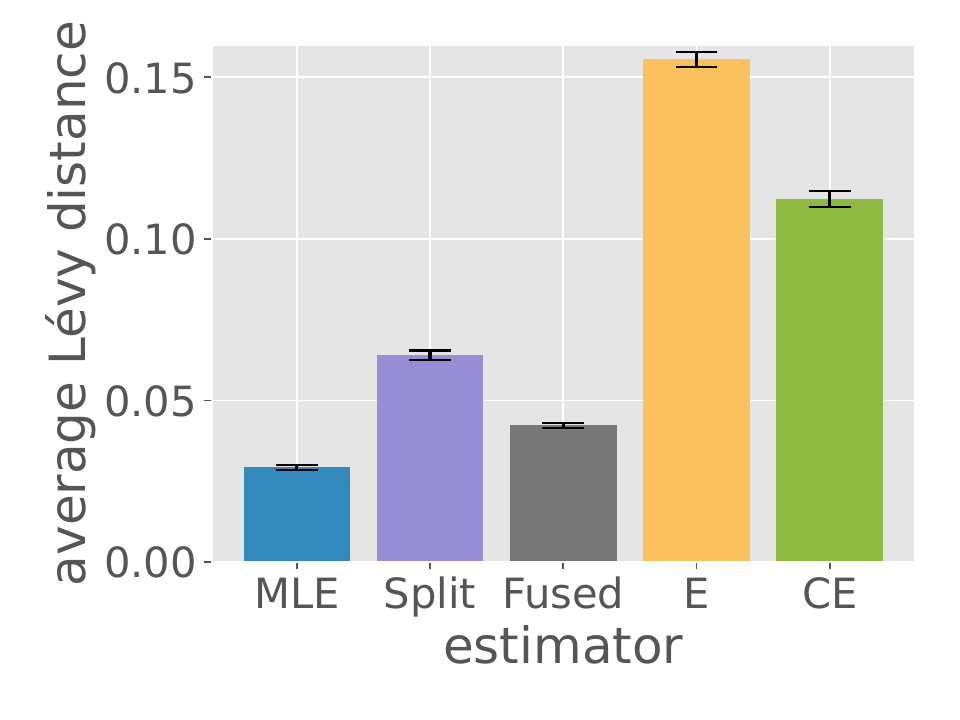}
    \\ \vspace{-0.8cm}
    \subcaption{For $n_0 = 10$, $n_1 = 10$.}
  \end{minipage}
  \begin{minipage}[b]{1.0\linewidth}
    \centering
    \includegraphics[width=\figWidth\textwidth]{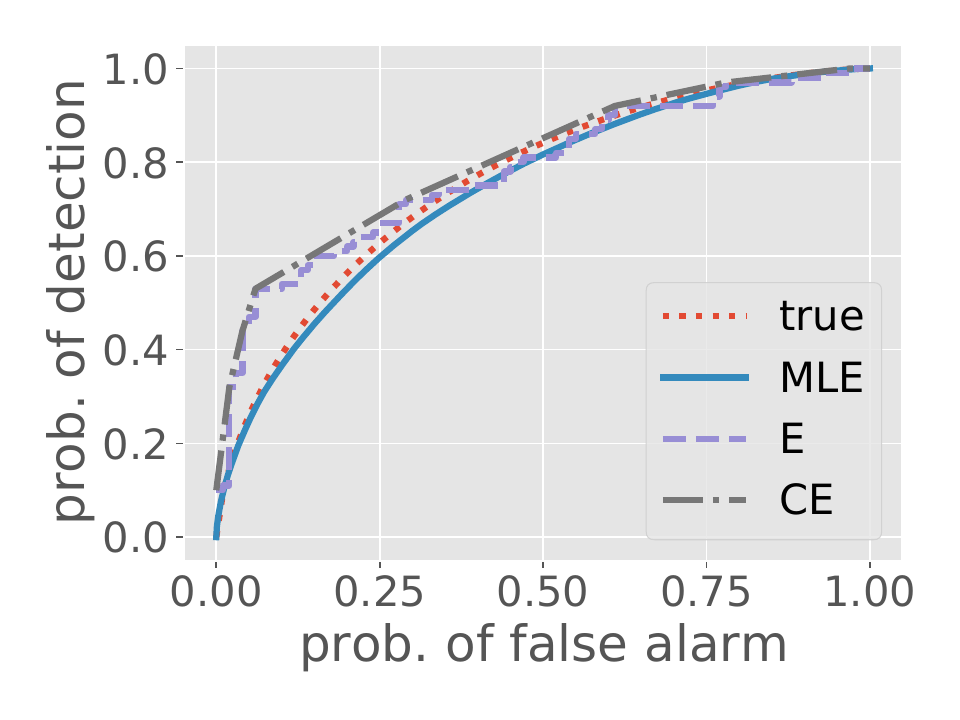}\includegraphics[width=\figWidth\textwidth]{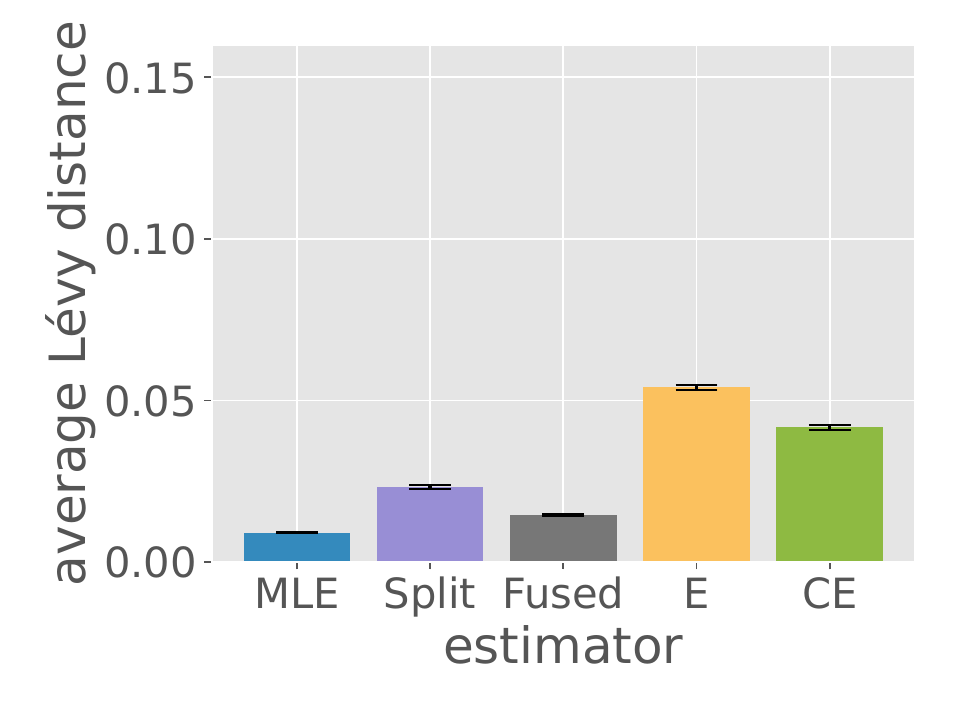}
    \\ \vspace{-0.8cm}
    \subcaption{For $n_0 = 100$, $n_1 = 100$.}
  \end{minipage}
  \begin{minipage}[b]{1.0\linewidth}
    \centering
    \includegraphics[width=\figWidth\textwidth]{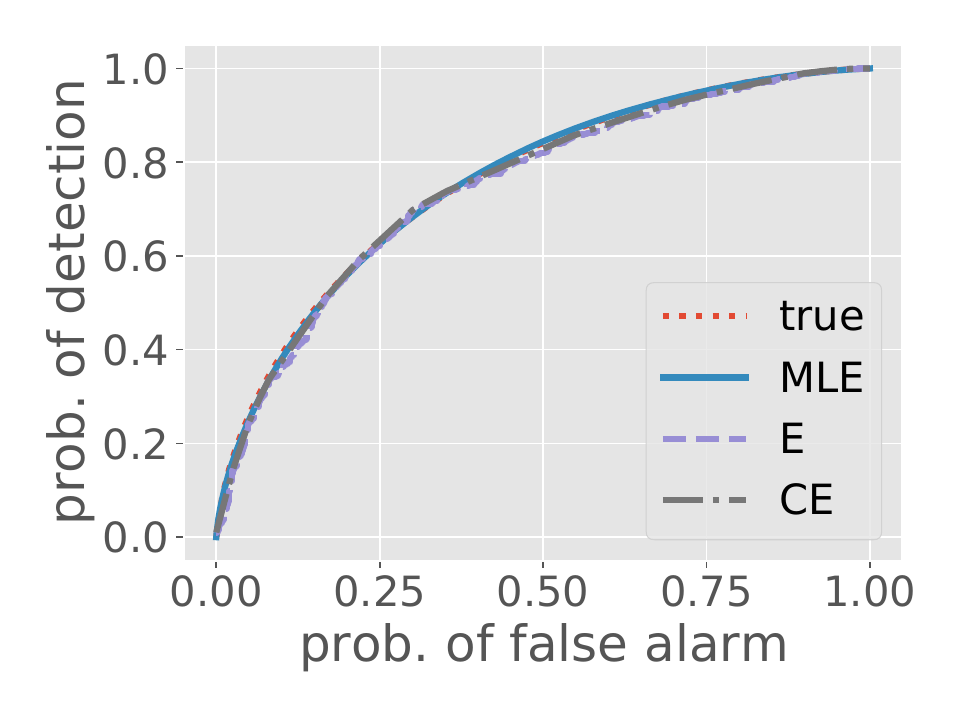}\includegraphics[width=\figWidth\textwidth]{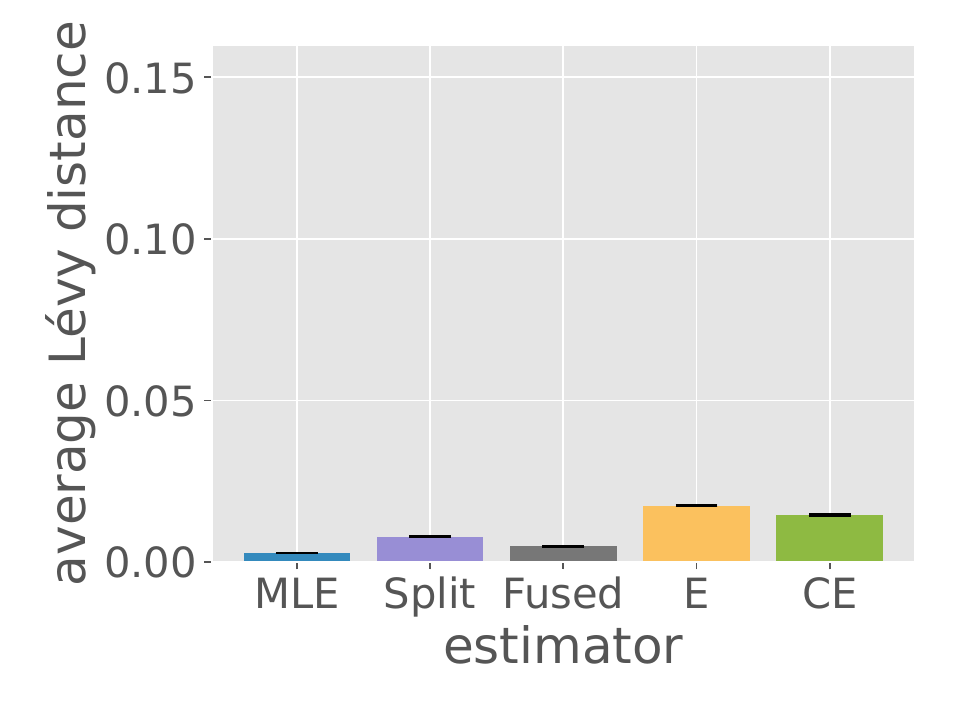}
    \\ \vspace{-0.8cm}
    \subcaption{For $n_0 = 1000$, $n_1 = 1000$.}
  \end{minipage}
  \caption{Sample instances and average errors for $\mu = 1$.}
  \label{fig:roc}
\end{figure}

\begin{figure}[thbp]
  \begin{minipage}[b]{1.0\linewidth}
    \centering
    \includegraphics[width=\figWidth\textwidth]{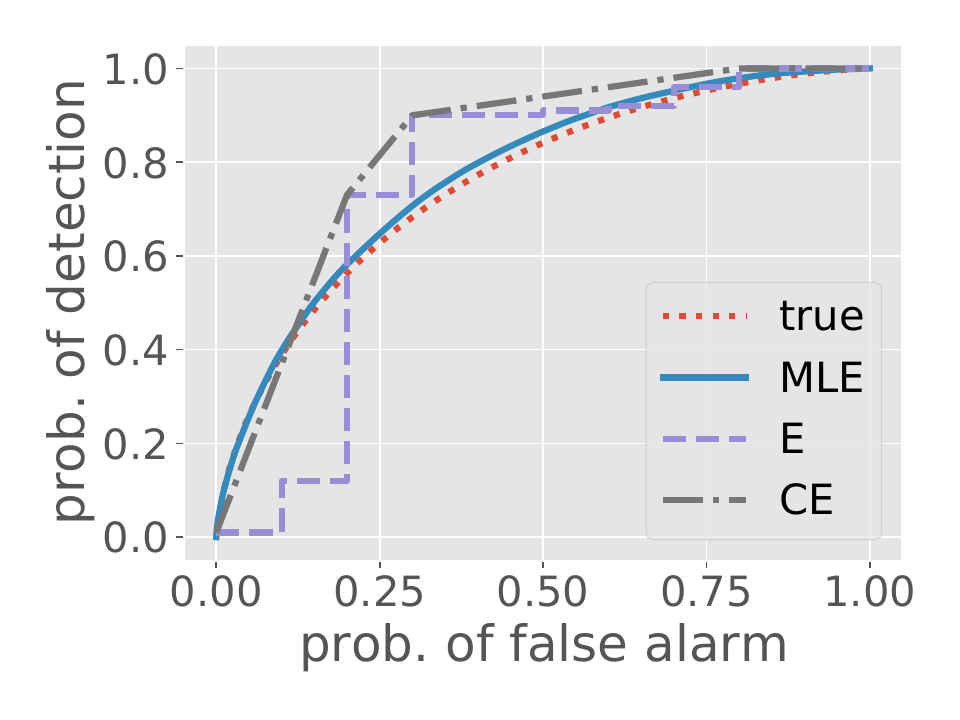}\includegraphics[width=\figWidth\textwidth]{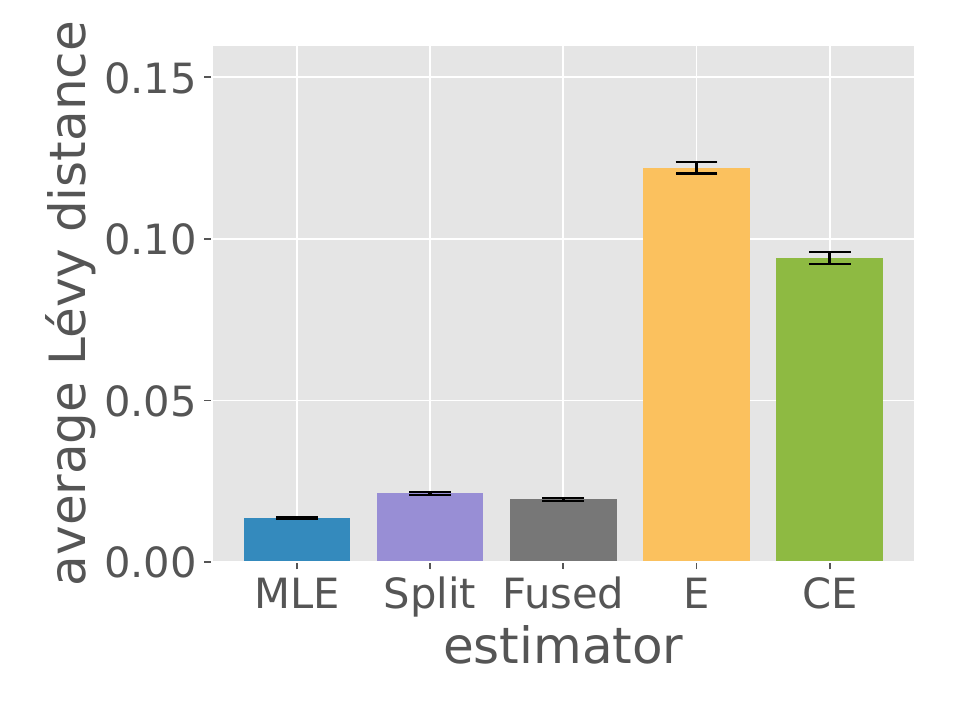}
    \\ \vspace{-0.8cm} \subcaption{For $n_0 = 10$, $n_1 = 100$.}
    \label{fig:roc4}
  \end{minipage}
  \begin{minipage}[b]{1.0\linewidth}
    \centering
    \includegraphics[width=\figWidth\textwidth]{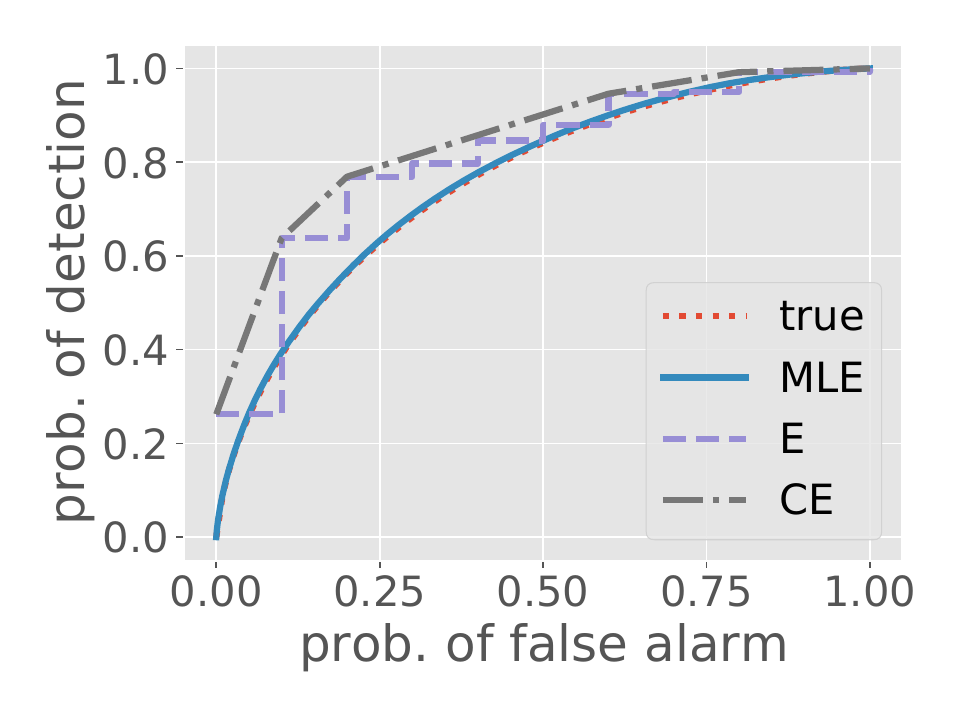}\includegraphics[width=\figWidth\textwidth]{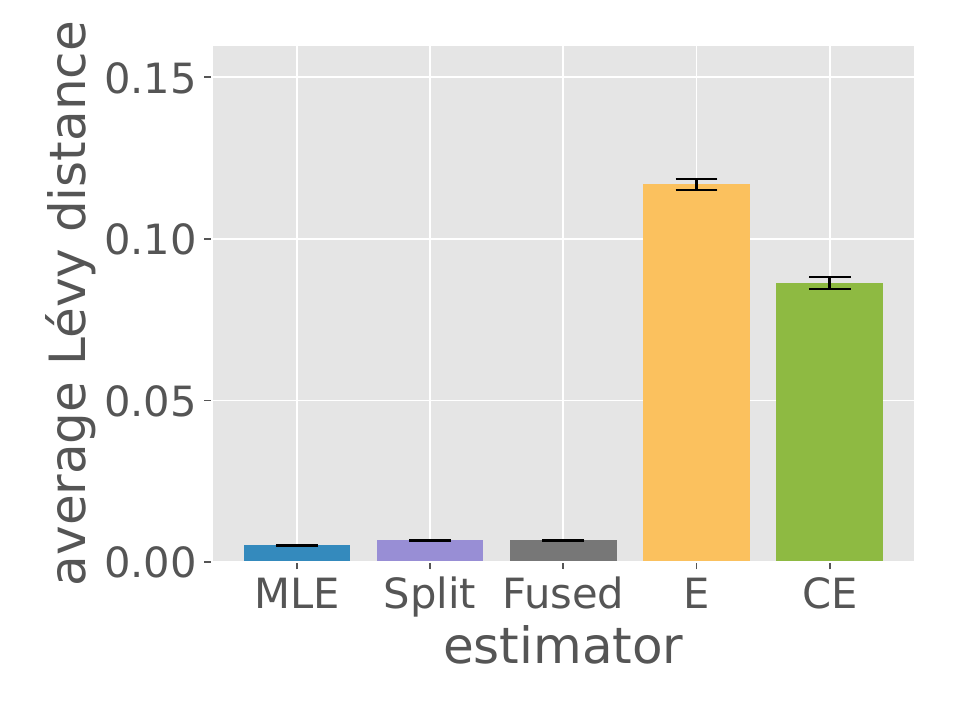}
    \\ \vspace{-0.8cm} \subcaption{For $n_0 = 10$, $n_1 = 1000$.}
    \label{fig:roc5}
  \end{minipage}
  \begin{minipage}[b]{1.0\linewidth}
    \centering
    \includegraphics[width=\figWidth\textwidth]{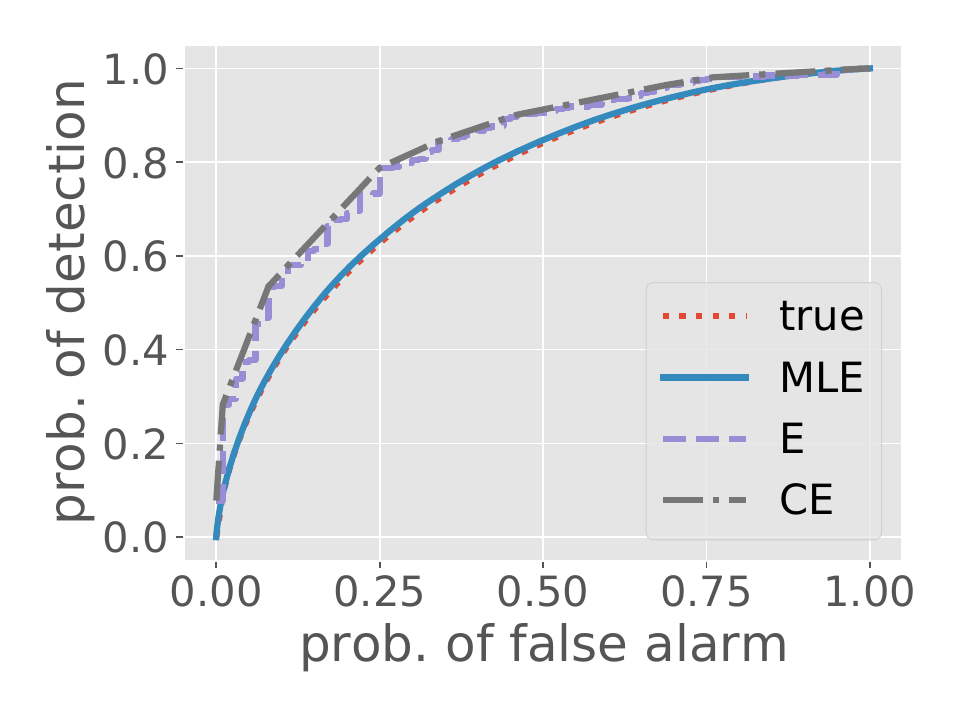}\includegraphics[width=\figWidth\textwidth]{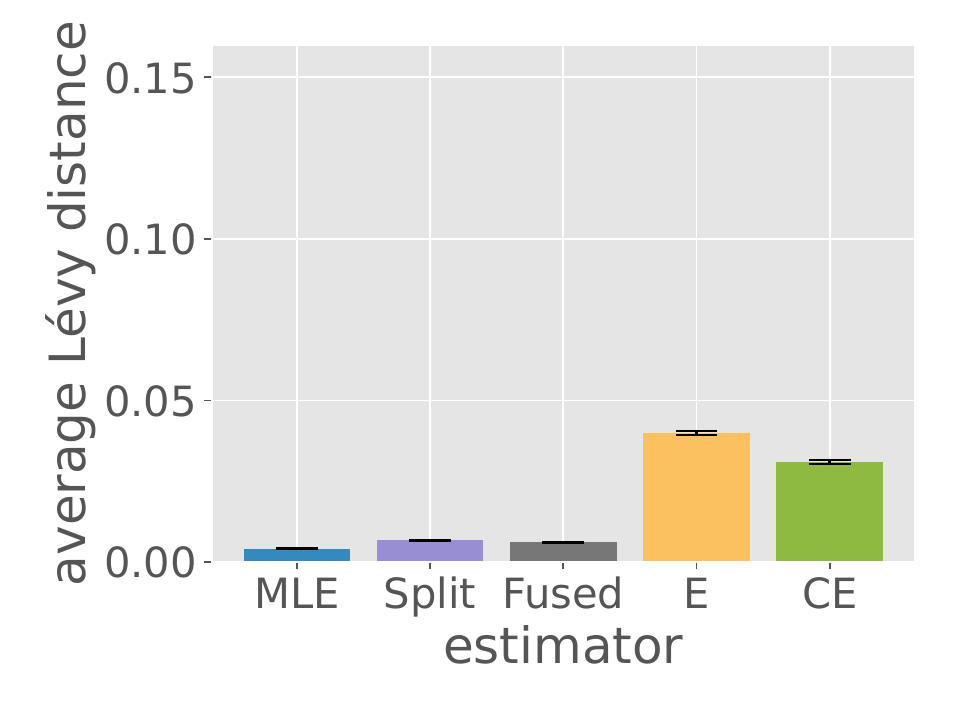}
    \\ \vspace{-0.8cm} \subcaption{For $n_0 = 100$, $n_1 = 1000$.}
    \label{fig:roc6}
  \end{minipage}
  \begin{minipage}[b]{1.0\linewidth}
    \centering
    \includegraphics[width=\figWidth\textwidth]{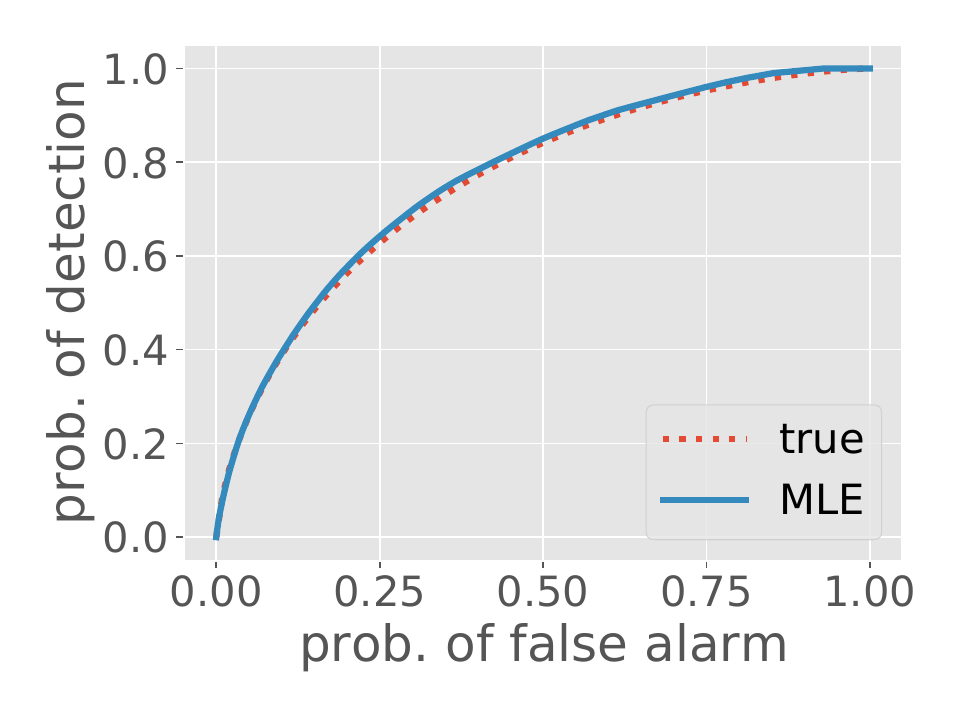}\includegraphics[width=\figWidth\textwidth]{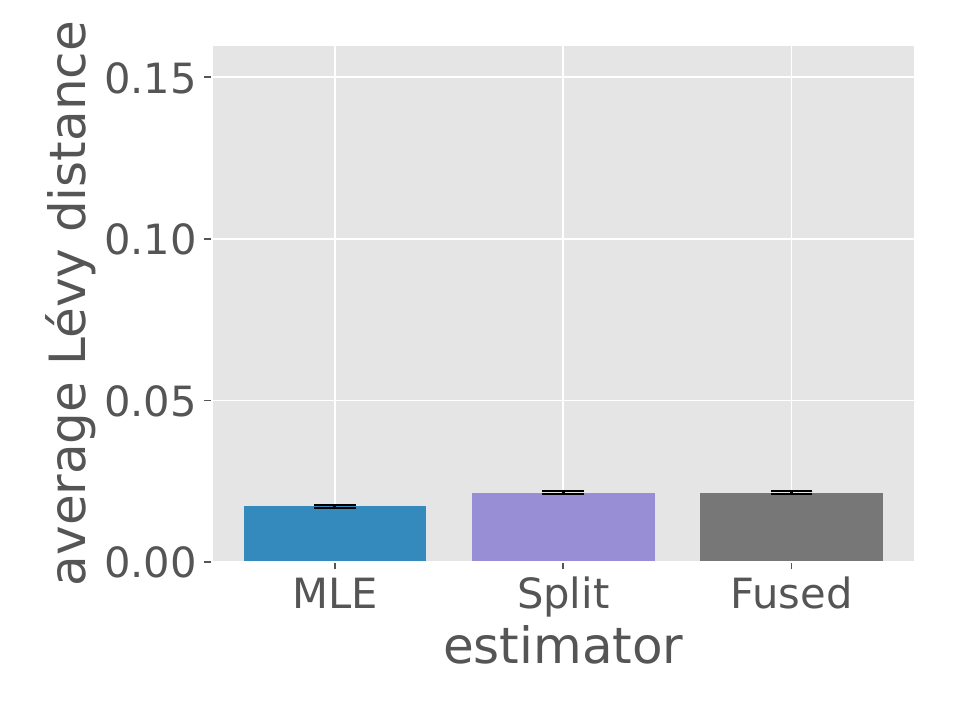}
    \\ \vspace{-0.8cm} \subcaption{For $n_0 = 0$, $n_1 = 100$.}
    \label{fig:roc7}
  \end{minipage}
  \caption{Sample instances and average errors for $\mu = 1$ (continued).}
  \label{fig:roc_cont}
\end{figure}

Simulation results for the ROC estimators with $\mu = 1$ are shown in Figs.~\ref{fig:roc} and \ref{fig:roc_cont} with various numbers of observations under the two hypotheses, $(n_0, n_1)$.  For each pair of $(n_0, n_1)$ two figures are shown.  The left figure shows samples of three of the estimators and the true ROC curve for a single sample instance of $n_0 + n_1$ likelihood ratio observations. (The split and fused estimators are not shown -- they are very close to the ML estimator.)  The right figure shows the average \levy\ distances of the estimators over $M = 500$ such sample instances with error bars (i.e., plus or minus sample standard deviations divided by $\sqrt M$).
The simulation code can be found at \cite{Kang22}.

The two empirical estimators have similar performance, while CE outperforms E slightly in terms of the average \levy\ distance.  Note $\ROCCE$, as the least concave majorant of $\ROCE$, could be biased toward higher probability of detection as evidenced by the sample instances.

It can be seen that the ML estimator (MLE) achieves much smaller average \levy\ distance than E or CE.  The difference is more pronounced when the number of observations under one hypothesis is significantly smaller than under the other, as seen in Figs.~\ref{fig:roc4}–\ref{fig:roc6}.  This is because E and CE calculate the empirical distributions based on the likelihood ratio observations under the two hypotheses separately before combining the empirical distributions into an estimated ROC curve.  As a result, having very few samples under either hypothesis results in errors in estimating the ROC curve regardless of how accurate the estimated distribution under the other hypothesis is.  In contrast, every observation contributes to the joint estimation of the pair of distributions in ML, so the ROC curve can be accurately estimated even when there are very few samples from one hypothesis.  The ML estimator and the split and fused variants work even if all samples are generated from the same hypothesis (see Fig.~\ref{fig:roc7}), while E and CE do not work because one of the distributions cannot be estimated at all.

Empirically, the ML estimator has a slightly smaller average error than the split or fused estimators and the difference between the split and fused estimators is even smaller, with the fused estimator being very slightly more accurate than the split estimator.

\begin{figure}[t]
  \begin{minipage}[t]{0.49\linewidth}
    \centering
    \includegraphics[scale=0.5]{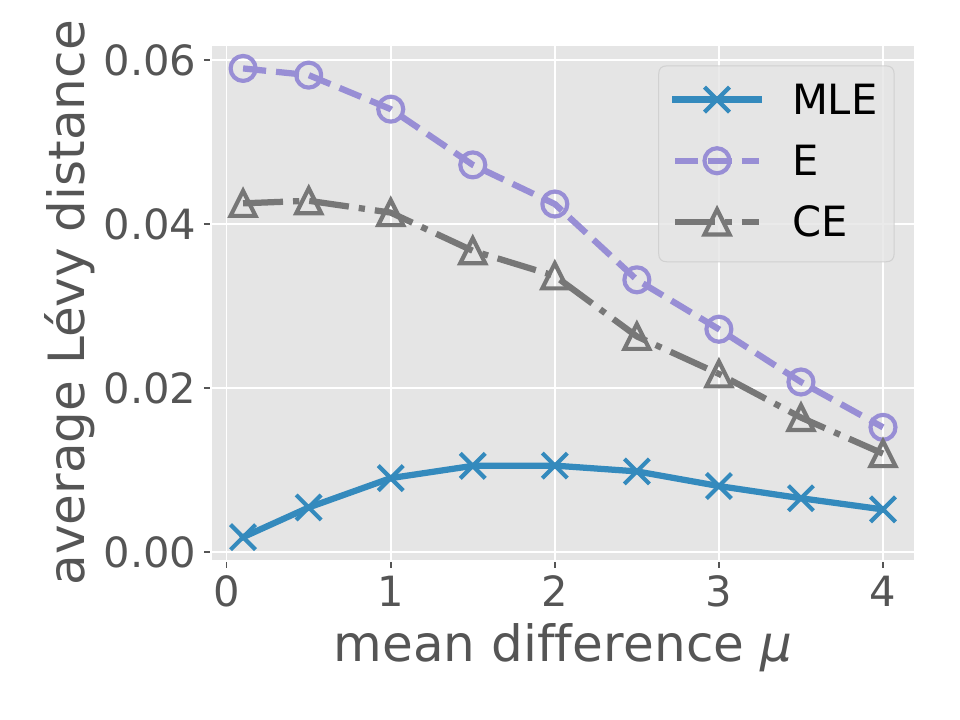}
    \\ \vspace{-0.5cm} \subcaption{For $n_0 = n_1 = 100$.}
    \label{fig:vary-mu}
  \end{minipage}
  \begin{minipage}[t]{0.49\linewidth}
    \centering
    \includegraphics[scale=0.5]{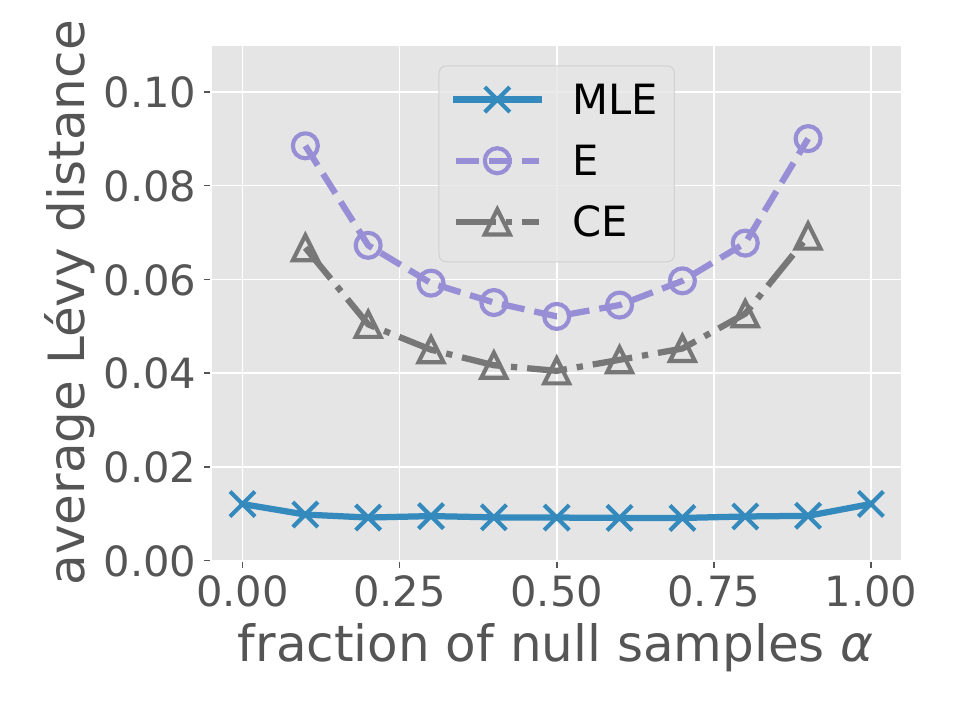}
    \\ \vspace{-0.5cm} \subcaption{For $\mu = 1$ and
      $n_0 + n_1 = 200$.}
    \label{fig:vary-alpha}
  \end{minipage}
  \vspace{-0.2cm}
  \caption{Average \levy\ distance for varying $\mu$ (left) or $\alpha$ (right).}
  \label{fig:vary-mu-alpha}
  \vspace{-0.6cm}
\end{figure}

Sensitivity of the performance of the estimators to the mean difference $\mu$ and to the sample composition $\alpha = n_1 / (n_0 + n_1)$ are shown in Fig~\ref{fig:vary-mu-alpha}, again averaged over $M = 500$ instances.
In the subfigure on the left, different values of $\mu$ are used for $n_0 = n_1 = 100$. In the subfigure on the right, different values of $\alpha$ are used for $\mu = 1$ and a fixed total number of samples $n_0 + n_1 = 200$.  In both cases, ML outperforms E and CE consistently and is less sensitive to $\mu$ and $\alpha$.

\begin{figure}[htb]
  \begin{minipage}[t]{0.5\linewidth}
    \centering
    \includegraphics[scale=0.6]{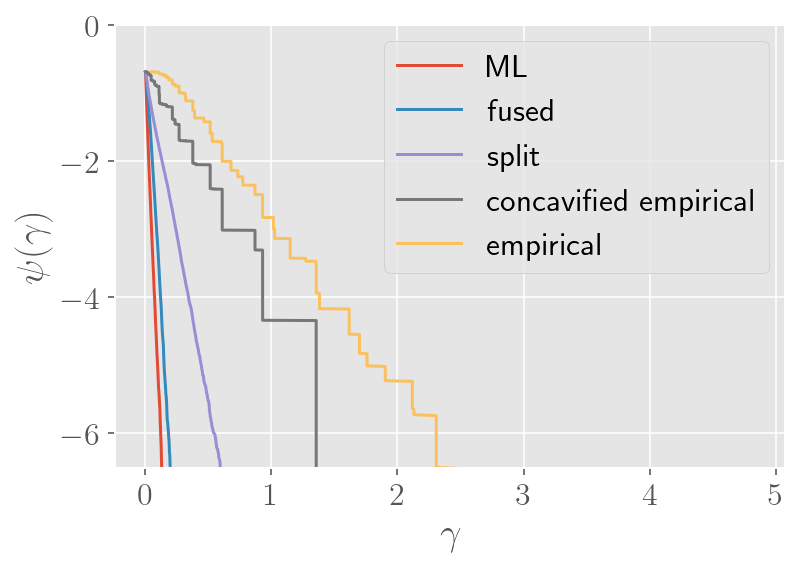}
    \\ \vspace{-0.5cm} \subcaption{$n=20, \alpha = 0.5.$}
  \end{minipage}
  \begin{minipage}[t]{0.5\linewidth}
    \centering
    \includegraphics[scale=0.6]{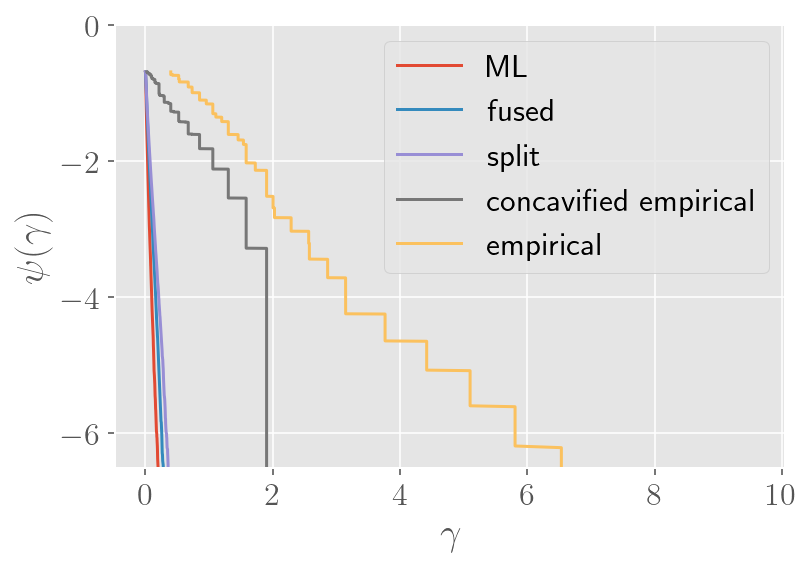}
    \\ \vspace{-0.5cm} \subcaption{$n=20, \alpha = 0.1$.}
  \end{minipage}
    \begin{minipage}[t]{0.5\linewidth}
    \centering
    \includegraphics[scale=0.6]{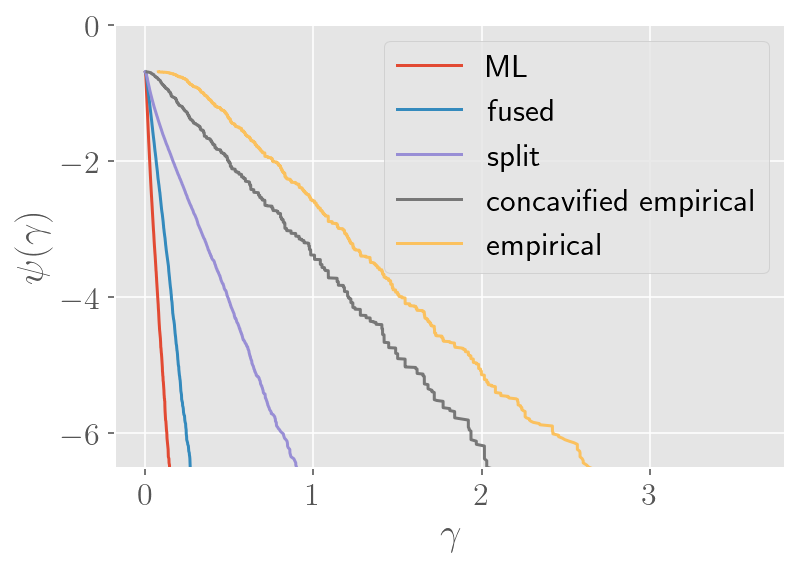}
    \\ \vspace{-0.5cm} \subcaption{$n=100, \alpha = 0.5.$}
  \end{minipage}
  \begin{minipage}[t]{0.5\linewidth}
    \centering
    \includegraphics[scale=0.6]{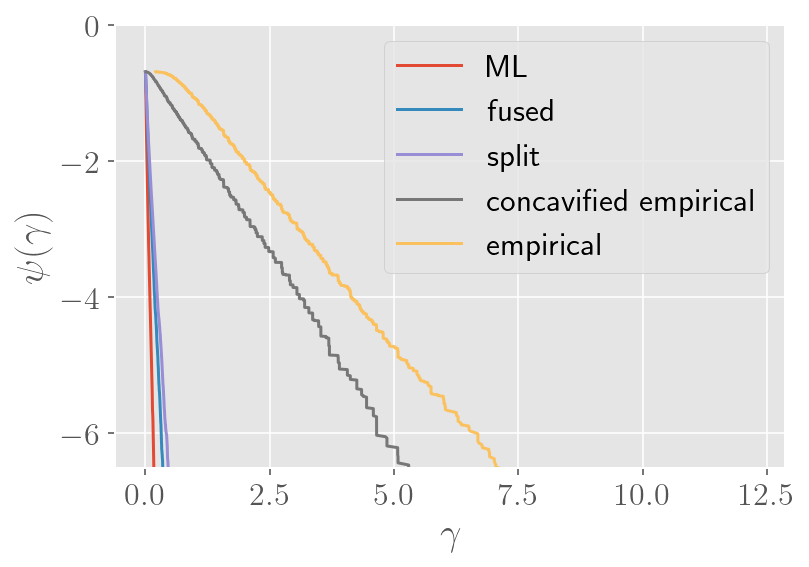}
    \\ \vspace{-0.5cm} \subcaption{$n=100, \alpha = 0.1$.}
  \end{minipage}
    \begin{minipage}[t]{0.5\linewidth}
    \centering
    \includegraphics[scale=0.6]{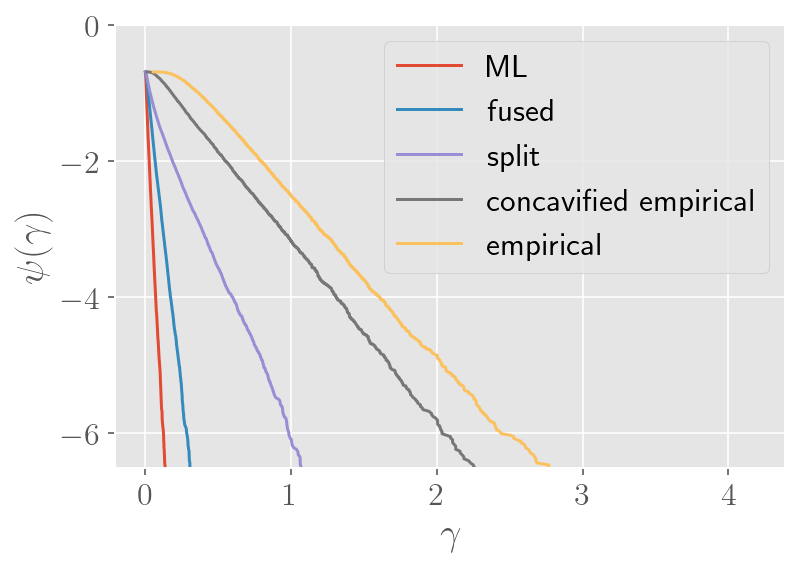}
    \\ \vspace{-0.5cm} \subcaption{$n=500, \alpha = 0.5.$}
  \end{minipage}
  \begin{minipage}[t]{0.5\linewidth}
    \centering
    \includegraphics[scale=0.6]{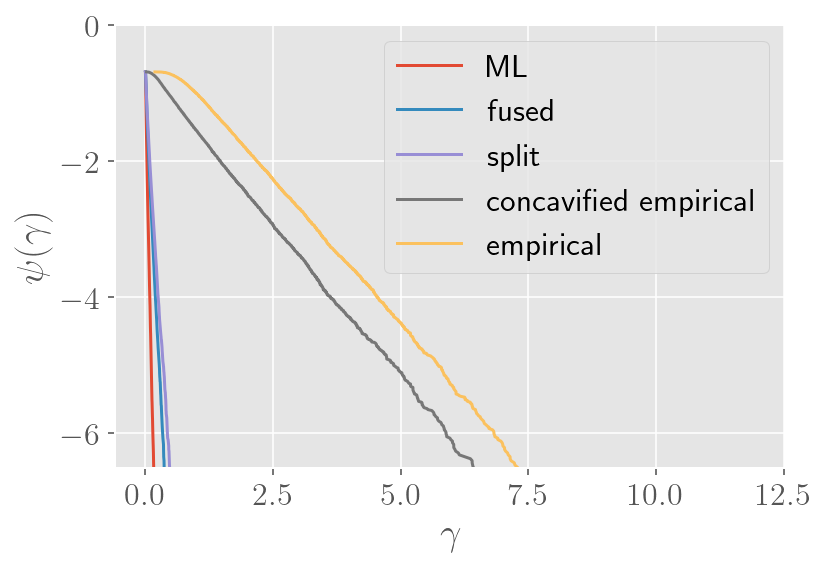}
    \\ \vspace{-0.5cm} \subcaption{$n=500, \alpha = 0.1$.}
  \end{minipage}
  \vspace{-0.2cm}
  \caption{Estimates of $\psi_n(\gamma)$ vs. $\gamma$, where $\psi_n(\gamma)$ is defined in \eqref{eq:log_bnd}, for the various estimators for $n\in\{20,100,500\}$ and $\alpha\in \{0.5,0.1\}$ for the binormal BHT with $\mu = 1.$   The plots are based on the \levy\ distances of M=10,000 sample estimates of the ROC for each estimator.}
  \label{fig:tail_plots}
  \vspace{-0.6cm}
\end{figure}

We turn to numerical investigation of the {\em distribution} of the \levy\ distance of the estimators from the true ROC. The bounds on tail probabilities of the  \levy\ distance $L = L(\hat{\ROC},\ROC)$  for the estimators in Proposition \ref{prop:finite_sample_bnds} have the form
\begin{align}  \label{eq:exp_bnd_various}
  \Pr\left\{L \geq \delta\right\} \le  2 \exp(-nc\delta^2)
\end{align}
for any $\delta > 0$ and integer $n \geq 1$ for some constant $c$ depending on $\alpha$.  Here, $n$ is the number of likelihood ratio samples used for each instance of $\hat{\ROC}$.  The bound in Proposition~\ref{prop:dkw-empirical} is similar.  Equivalently, letting $\delta = \sqrt{\frac \gamma n}$ and taking the logarithm on each side of \eqref{eq:exp_bnd_various} yields
\begin{align}
  \label{eq:log_bnd}
  \psi_n(\gamma) \stackrel{\triangle}{=} \log  \left( \frac 1 2 \Pr\left\{L \geq \sqrt{\frac \gamma n} \right\} \right) \leq -c\gamma
\end{align}
for any $\gamma > 0.$  While each bound in Proposition \ref{prop:finite_sample_bnds}  provides a value of $c$ depending only on $\alpha$, the proof techniques might not yield the best possible value of $c$ and therefore might not correctly rank the estimators by their accuracy.  To investigate what may be the largest valid choice of $c$ for a given estimator and value of $\alpha,$ we plot an estimate of $\psi_n$ for each of the estimators for $n\in \{20,100, 500\}$, based on Monte Carlo simulation and try to identify a slope $-c$ for each one such that \eqref{eq:log_bnd} holds.  If $L_1, \dots, L_M$ are $M$ independent samples of $L$ we use the empirical distribution of these samples to get
\begin{align*}
\psi_n(\gamma)  \approx \log \left( \frac{ \bigg| \left\{j: L_j \geq \sqrt{\frac \gamma n} \right\} \bigg| } {2M} \right).
\end{align*}
Thus, if we sort the samples so $L_1 < \dots < L_M$, we want
\begin{align*}
  \gamma_i \stackrel{\triangle}{=} nL_i^2 \mapsto \log \frac{M-i+1}{2M},
\end{align*}
because $M-i+1$ of the samples are greater than or equal to $L_i$ (assuming no ties).  So we plot the pairs $\left( nL_i^2 , \log \frac{M-i+1}{2M} \right)$ for $i\in [M]$ to accurately approximate the graph of $\psi_n.$
Such plots are shown in Fig.~\ref{fig:tail_plots} for $n\in \{20,100, 500\}$ and $\alpha\in \{0.5,0.1\}$ for the binormal BHT problem.  The curves are nearly straight lines except those for the empirical and concavified empirical estimators when $n=20$.  (Those estimators perform poorly for such a small number of observations and the fact the distribution of $\hat{\ROC}$ is discrete for them is evident.)  Note that the downward slopes are considerably larger for the ML, fused, and split estimators in contrast to the slopes for the empirical and concavified empirical estimators.

The following are examples of statements that can be made based on Fig. \ref{fig:tail_plots} for the binormal BHT.  Since $\psi_{n}(0.16) < -6$ for $n\in \{20,100, 500\}$ and $\alpha\in \{0.5,0.1\}$ for the ML estimator, we conclude the following for such $n$ and $\alpha$.  Based on $n$ likelihood ratio samples, the ML estimator achieves $\Pr\left\{L(\ROC, \ROCML)\leq \delta \right\} \geq 1-e^{-6} \geq 0.9975$ with $\delta =\sqrt{\frac{0.16} n} = 0.09, 0.04,$ or $0.02$ for $n=20,100,$ or $500$, respectively.  In contrast, the following representative statement we can make for the concavified empirical estimator is considerably weaker.  For the concavified empirical estimator, $\psi_{n}(1) \leq -3$ for $n\in \{20,100, 500\}$ and $\alpha=0.5$.  Therefore, based on $n$ likelihood ratio samples with $\alpha = 0.5$, the concavified empirical estimator achieves $\Pr\left\{L(\ROC, \ROCCE)\leq \delta \right\} \geq 1-e^{-3} \geq 0.95$ with $\delta =\sqrt{\frac 1 n} = 0.23, 0.1,$ or $0.045$ for $n=20,100,$ or $500$, respectively.

We observe from the figures that the functions $\psi_n$ have a very small dependence on $n$ so that we can translate the negative slopes into numbers of likelihood ratio samples needed for a given accuracy because $n$ and $c$ appear in the right hand side of \eqref{eq:exp_bnd_various} only through their product, $nc.$   Specifically, for $\alpha = 0.5$, the negative slope for the ML estimator is $c\approx 30$ and for the concavified empirical estimator (for $n\in\{100,500\}$) is $c\approx 2.5$. (The value $c=30$ is 15 times larger than the largest value in Fig. \ref{fig:c_vs_alpha}. And the value $c=2.5$ for the concavified empirical estimator is larger than the guarantee of $c=1$ for the empirical estimators shown in Fig. \ref{fig:c_vs_alpha}.)   The observed slopes imply that for the same accuracy, $\frac{30} {2.5}\approx 12$ times as many likelihood ratio observations are needed by the concavified empirical estimator as by the ML estimator, for this binormal BHT.   Comparing the plots in Fig. \ref{fig:tail_plots} for $\alpha = 0.1$ to those for $\alpha = 0.5$ shows that the slopes for the first two estimators are nearly the same as for both $\alpha$ values while the concavified empirical estimator has a much smaller magnitude slope (about -1) for $\alpha = 0.1$ suggesting $c\approx 1$ for that estimator for $\alpha = 0.1.$  This implies that for the same accuracy 30 times as many likelihood ratio observations are needed by the concavified empirical estimator as by the ML estimator for this binormal example with $\alpha = 0.1$ .

\begin{figure}[thb]
  \begin{minipage}[t]{0.5\linewidth}
    \centering
    \includegraphics[scale=0.6]{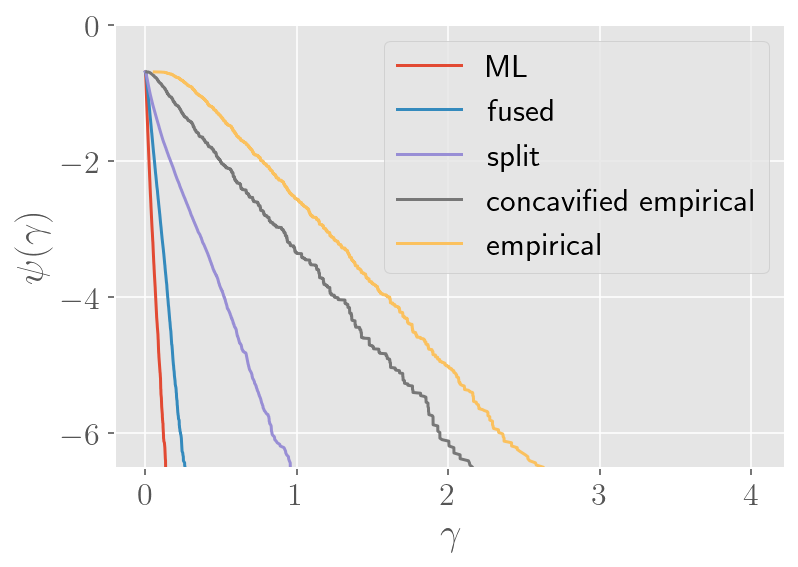}
    \\ \vspace{-0.5cm} \subcaption{$n=100, \alpha = 0.5.$}
  \end{minipage}
  \begin{minipage}[t]{0.5\linewidth}
    \centering
    \includegraphics[scale=0.6]{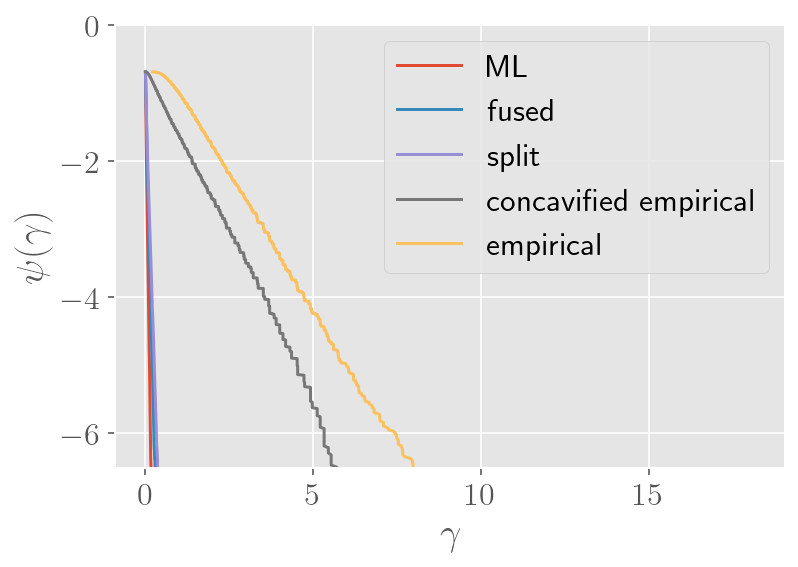}
    \\ \vspace{-0.5cm} \subcaption{$n=100, \alpha = 0.1$.}
  \end{minipage}
  \vspace{-0.2cm}
  \caption{Estimates of $\psi_{100}(\gamma)$ vs. $\gamma$ for the various estimators for the BHT such that under $H_0$ the likelihood ratio has the exponential distribution with mean one.      The plots are based on M=10,000 sample estimates of the ROC for each estimator.}
  \label{fig:tail_plots_gamma}
  \vspace{-0.6cm}
\end{figure}

The same calculations used to produce Figure \ref{fig:tail_plots} were used to produce Figure \ref{fig:tail_plots_gamma} for the BHT problem with $f_0(r) = e^{-r}$ for $r>0$ and $f_1(r)=re^{-r}.$ The distribution of the likelihood ratio under $H_1$ is the gamma distribution with shape parameter 2 and if $F_1$ denotes the corresponding CDF then the ROC curve is given by $p_{det} = F^c_2(-\log(p_{fa})).$  The performance of the estimators for this BHT is very close to their performance for the binormal BHT discussed above.

To conclude this section, we comment on the relative performance of the estimators for first and second halves of this section.
The overall relative performance of the estimators is the same for comparison of mean \levy\ distance and distribution of \levy\ distance, with the ML estimator being the most accurate, followed closely by the fused and split estimators.  All three are significantly more accurate than the two empirical estimators, especially when $\alpha$ is not close to 0.5.
It is also striking that the ML estimator and its variants are considerably more accurate than the finite sample size performance guarantees of Proposition \ref{prop:finite_sample_bnds}.  Of course those bounds hold for {\em any} BHT while in this section we focus on the binormal BHT and in Fig. \ref{fig:tail_plots_gamma} we touched on one other BHT.

\clearpage
\section{Conclusions and future directions}
\label{sec:conclusions}
The qualitative differences between the concavified empirical estimator $\ROCCE$ and the ML estimator $\ROCML$ are striking. Only the rank ordering of the samples is used by the concavified empirical estimator--not the numerical values.  So it is important to track which samples are generated with which distribution.  The ML estimator does not depend on which samples were generated with which distribution and exact numerical values are used.

The simulations in Section~\ref{sec:simulations} investigating the distribution of the \levy\ distance of the ML, fused, and split estimators show them to be much more accurate than the empirical estimators, for the binormal BHT problem.  It would be interesting to find tighter performance guarantees than those we have found, possibly with some mild conditions on the BHT, that come close to matching the performance differences observed in the simulations.  The simulations suggest that the differences in performance could come down to different values of the constant $c$, suggesting a constant factor (in $n$) relationship between the number of samples needed by one estimator to achieve the same performance as another estimator.  While the difference in constants $c$ might turn out to be large (on the order of ten or more, depending on $\alpha$), the simulations suggest there is not a superlinear relationship.  Therefore, the difference in performance might be most significant in applications where the number of samples $n$ is moderate, as in the simulations, and in that case difficult to quantify in a theoretical way.

A BHT is equivalent to a binary input channel.  Work of Blackwell and others working on the comparison of experiments has led to canonical channel descriptions that are equivalent to the ROC curve, such as the Blackwell measure.  The Blackwell measure is the distribution of the posterior probability that hypothesis $H_0$ is true for equal prior probabilities $1/2$ for the hypotheses.  See \cite{GoelaRaginsky20} and references therein.  It may be of interest to explore estimation of various canonical channel descriptions besides the ROC under various metrics.

\section*{Acknowledgments}
This material is based upon work supported by the National Science Foundation under Grant No.\ CCF 19-00636.

\appendices
\section{Relation of $F_0$ and $F_1$}
\label{app:relation}
Let $P_k$ and $g_k$ denote the probability distribution and the probability density function with respect to some reference measure $\mu$ of the observation $X$ in a measurable space $(\mathcal X, \Sigma)$ under hypothesis $H_k$ for $k = 0, 1$.  In other words, $P_k(A)=\int_Ag_k(x)\mu(\d x)$ for any $A\in\Sigma$.  Let $\rho\colon\mathcal X\to\extReals\defeq\reals\cup\{\infty\}$ be defined by
\[\rho(x) =
  \begin{cases} \frac{g_1(x)}{g_0(x)} & \text{if }g_0(x) > 0,\\
    \infty & \text{if }g_0(x) = 0.
  \end{cases}
\]
Then $\rho$ is a Borel measurable function denoting the likelihood ratio given an observation.  The probability distribution of the extended random variable $R = \rho(X)$ under $H_k$ is the push-forward of the measure $P_k$ induced by the function $\rho$ for $k = 0, 1$, denoted by $\nu_k$.  The probability distribution $\nu_k$ restricted to $\reals$ is also the unique Borel measure (known as the Lebesgue–Stieltjes (L–S) measure corresponding to $F_k$, the CDF of $R$) on $[0,\infty)$ such that $\nu_k([0, \tau]) = F_k(\tau)$ for all $\tau \in [0,\infty)$.

Throughout this paper, integrals of the form $\int h(r) \d F(r)$ are understood to be Lebesgue–Stieltjes integrals (for the extended real numbers).  That is,
\begin{align*}
  \int_{\extReals} h(r) \d F(r) \defeq \int_{\extReals} h(r) \nu_F(\d r),
\end{align*}
for any Borel measurable function $h$.

\begin{prop}
  For any Borel subset $A$ of $\reals$,
  \[\nu_1(A) = \int_Ar\nu_0(\d r).\]
  In other words, when restricted to the Borel sets in $\reals$, $\nu_1$ is absolutely continuous with respect to $\nu_0$, and the Radon–Nikodym derivative is the identity function almost everywhere with respect to $\nu_0$.
\end{prop}
\begin{IEEEproof}
  By the change-of-variables formula for push-forward measures, for any Borel set $A$ in $\reals$,
  \begin{align*}
    \nu_1(A) & = \int_{\extReals}\indicator A(r)\nu_1(\d r)\\
             & = \int_{\mathcal X}\indicator A(\rho(x)) P_1(\d x)\\
             & = \int_{\mathcal X}\indicator A(\rho(x)) g_1(x)\mu(\d x)\\
             & = \int_{\mathcal X}\indicator A(\rho(x)) \rho(x)g_0(x)\mu(\d x)\\
             & = \int_{\mathcal X}\indicator A(\rho(x))\rho(x) P_0(\d x)\\
             & = \int_{\extReals}\indicator A(r)r \nu_0(\d r)\\
             & = \int_Ar\nu_0(\d r),
  \end{align*}
  implying the proposition.
\end{IEEEproof}

\section{Proofs for Section~\ref{sec:formulation} -- Preliminaries}

\begin{IEEEproof}[Proof of Proposition~\ref{prop:equivalence of Fs and ROC}]
  The function $F_0$ determines $F_1$ by $F_1(\tau)=\int_{[0, \tau]} r \d F_0(r)$ for $\tau\in[0, \infty)$.  Conversely, $F_1$ determines $F_0$ by $F_0^c(\tau)=\int_{(\tau, \infty)} \frac 1 r \d F_1(r)$ for $\tau\in[0, \infty)$.  So either one of $F_0$ or $F_1$ determines the other, and hence also determines $\ROC$ as described in Section~\ref{sec:model}.  To complete the proof it suffices to show that $\ROC$ determines $F_0$.  The function $\ROC$ is concave so it has a right-hand derivative on $[0, 1)$ which we denote by $\ROC'$, with the understanding that $\ROC'(0)\in[1, \infty]$ and the convention that $\ROC'(1) = 0$.  Then we have $F^c_0(\tau) = \min\set{p\in[0, 1]}{\ROC'(p) \le \tau}$ for $\tau\in[0, \infty)$.
\end{IEEEproof}

\begin{IEEEproof}[Proof of Lemma~\ref{lemma:BHT_metric_bound}]
  Let the right-hand side of \eqref{eq:bound_on_L_dist} be denoted by $\epsilon$.  Note that
  \begin{align}
    \epsilon = \sup_{\tau \in (0,\infty), \eta\in [0,1]} \max\{ & |F^c_{a,0}(\tau,\eta)-F^c_{b,0}(\tau,\eta)|,\nonumber\\
                                                                &|F^c_{a,1}(\tau,\eta)-F^c_{b,1}(\tau,\eta)| \},   \label{eq:bound_on_L_dist_long}
  \end{align}
  because for $\tau$ fixed, the right-hand side of \eqref{eq:bound_on_L_dist_long} is the maximum of a convex function of $\eta$ and the value at $\eta=0$ and $\eta=1$ is obtained by the right-hand side of \eqref{eq:bound_on_L_dist} at $\tau-$ and $\tau,$ respectively.  We appeal to the geometric interpretation of $L(A,B)$.  Consider any point $(p, B(p))$ on the graph of $B$.  It is equal to $(F^c_{b, 0}(\tau, \eta), F^c_{b, 1}(\tau, \eta))$ for some choice of $(\tau, \eta)$.  Let $(p', A(p'))$ denote the point on the graph of $A$ for the same choice of $(\tau, \eta)$.  In other words, it is the point $(F^c_{a,0}(\tau,\eta), F^c_{a,1}(\tau,\eta))$.  Then $(p, B(p))$ can be reached from $(p', A(p'))$ by moving horizontally at most $\epsilon$ and moving vertically at most $\epsilon$.  So $(p, B(p))$ is contained in the region bounded between the upper and lower shifts of the graph of $A$ as claimed.
\end{IEEEproof}

\section{Proof for Section~\ref{sec:empirical_estimator} - The Empirical Estimator}
\label{sec_using_DKW}
\begin{IEEEproof}[Proof of Proposition~\ref{prop:dkw-empirical}]
  Combining the DKW inequality \eqref{eq:DKWM} with Lemma~\ref{lemma:BHT_metric_bound} implies \eqref{eq:empirical_estimator_bnd}.  The consistency of $\ROCE$ follows from the Borel–Cantelli lemma and the fact the sum of the right-hand side of \eqref{eq:empirical_estimator_bnd} over $n$ is finite for any $\delta > 0$.

  The final inequality follows from the following observations: $\ROCCE(p) \geq \ROCE(p)$ for $p\in [0, 1]$, and if $\ROCE$ is less than or equal to the concave function $p\mapsto\ROC (p+\epsilon)+\epsilon,$ then so is $\ROCCE$, by the definition of least concave majorant.
\end{IEEEproof}

\section{Proofs for Section~\ref{sec:mle} -- the ML ROC estimator}

\subsection{Derivation of $\ROCML$}
\label{app:derivation_ML}

Proposition~\ref{prop:ML_estimator} and its corollary are proved in this section.

\begin{IEEEproof}[Proof of Proposition~\ref{prop:ML_estimator}]
  Given the binary sequence $\ind{I_i}{i\in [n]}$ and the likelihood ratio samples $R_1, \dots, R_n$, let $0 = v_0 < v_1 < v_2 < \dots < v_m < v_{m + 1} = \infty$ be the set of unique values of the samples, augmented by $v_0=0$ and $v_{m+1}=\infty$ even if $0$ and/or $\infty$ is not among the observed samples.  Let $(c_0^0, c_1^0, c_2^0, \dots, c_m^0)$ denote the multiplicities of the values from among $\ind{R_i}{I_i = 0}$ and let $(c_1^1, c_2^1, \dots, c_m^1, c_{m + 1}^1)$ denote the multiplicities of the values from among $\ind{R_i}{I_i = 1}$.

  Let $a_j = F_0(\{v_j\})$ for $0 \leq j \leq m$ and let $b = F_1(\{\infty\})$.  Thus $a_j$ is the probability mass at $v_j$ under hypothesis $H_0$ for $0\leq j \leq m$.  The corresponding probability mass at $v_j$ under hypothesis $H_1$ is $a_jv_j$ for $0\leq j \leq m$ and the probability mass at $v_{m+1}$ under hypothesis $H_1$ is $b$.

  The log-likelihood to be maximized is given by
  \begin{align*}
    \sum_{j = 0}^{m}c_j^0\log\pmfLike_j + \sum_{j = 1}^{m}c_j^1\log(\pmfLike_jv_j) + c_{m + 1}^1\log\pmfInfty,
  \end{align*}
  where $0\log 0$ is understood as $0$ and $\log 0$ is understood as negative infinity.  Equivalently, dropping the term $\sum_{j = 1}^{m}c_j^1\log(v_j)$ which does not depend on $F_0$ (or $F_1$ or $\ROC$), the ML estimator is to maximize
  \begin{align*}
    \sum_{j = 0}^{m}c_j\log\pmfLike_j + c_{m + 1}\log\pmfInfty,
  \end{align*}
  where $c_0 \defeq c_0^0$, $c_{m + 1} \defeq c_{m + 1}^1$ and $c_j \defeq c_j^0 + c_j^1$ for $1\le j\le m$.  In other words, $c_j$ is the total multiplicity of $v_j$ in all samples regardless of the hypothesis.

  The probabilities satisfy the constraint:
  \begin{align}
    \label{eq:prob_constraint}
    \sum_{j = 0}^{m}\pmfLike_j \leq 1 \text{ and } \sum_{j = 1}^{m}\pmfLike_jv_j + b \leq 1.
  \end{align}
  The inequalities in \eqref{eq:prob_constraint} both hold with equality if the distribution $F_0$ (or equivalently $F_1$) assigns probability one to the set $\{v_0, \dots, v_{m+1}\}$.  Otherwise, both inequalities are strict.  We claim and now prove that any ML estimator is such that both inequalities in \eqref{eq:prob_constraint} hold with equality.  It is true in the degenerate special case that $R_i \in \{0, \infty\}$ for all $i$ (equivalently, $m=0$), in which case an ML estimator is given by $\ROC(p)\equiv 1$, $F_0(0) = 1$ and $F_1(\{\infty\}) = 1$.  So we can assume $m\geq 1$ and there is a value $j_0$ (for example, $j_0 = 1$) such that $1 \leq j_0 \leq m$.  If $F_0$ does not assign probability one to $\{v_0, \dots, v_{m+1}\}$ then the same is true for $F_1$, so that strict inequality must hold in both constraints in \eqref{eq:prob_constraint}.  Then the probability mass from $F_0$ (and $F_1$) that is not on the set $\{v_0, \dots, v_{m+1}\}$ can be removed and mass can be added to $F_0$ at 0 and $v_{j_0}$ and to $F_1$ at $v_{j_0}$ and $\infty$ such that both constraints in \eqref{eq:prob_constraint} hold with equality and the likelihood is strictly increased.  This completes the proof of the claim.

  Therefore, any ML estimator is such that the distributions are supported on the set $\{v_0, \dots, v_{m+1}\}$ and the probabilities assigned to the points give an ML estimator if and only if they are solutions to the following convex optimization problem:
  \begin{align}
    \max_{\pmfLike \ge 0, \pmfInfty\ge 0} \quad& \sum_{j = 0}^{m}c_j\log\pmfLike_j + c_{m + 1}\log\pmfInfty\label{eq:mle}\\
    \text{s.t.} \quad& \sum_{j = 0}^{m}\pmfLike_j = 1 \text{ and } \sum_{j = 1}^{m}\pmfLike_jv_j + b = 1. \nonumber
  \end{align}
  Since the constraints are linear equality constraints and there exist feasible $(a,b)$ in the interior of the constraint set, the relaxed Slater constraint qualification condition is satisfied for \eqref{eq:mle}.  Therefore, there exists a solution and dual variables satisfying the KKT conditions (see Theorem 3.2.4 in \cite{Ben-TalNemirovski13}).  The Lagrangian is
  \begin{align*}
    L(\pmfLike, \pmfInfty, \mu, \lambda) & = \sum_{j = 0}^mc_j\log\pmfLike_j + c_{m + 1}\log\pmfInfty\\
                                         &\quad - \mu\left(\sum_{j = 0}^m\pmfLike_j - 1\right) - \lambda\left(\sum_{j = 1}^m\pmfLike_jv_j + b - 1\right).
  \end{align*}
  The KKT conditions on $(a,b,\mu,\lambda)$ are
  \begin{align*}
    & \pmfLike \ge 0, \pmfInfty\ge 0;\quad \sum_{j = 0}^{m}\pmfLike_j = 1;\quad \sum_{j = 1}^{m}\pmfLike_jv_j + b = 1;\\
    & \frac{\partial L}{\partial \pmfLike_0} \le 0;\quad a_0 \cdot \frac{\partial L}{\partial \pmfLike_0} = 0;\quad \frac{\partial L}{\partial a_j} =0 \text{ for } j\in [m];\\
    & \frac{\partial L}{\partial \pmfInfty} \le 0;\quad b \cdot \frac{\partial L}{\partial \pmfInfty} = 0,
  \end{align*}
  where
  \[\frac{\partial L}{\partial \pmfLike_0}(\pmfLike, \pmfInfty, \mu, \lambda) =
    \begin{cases}
      \frac{c_0}{\pmfLike_0} - \mu & \text{if }c_0 > 0,\\
      - \mu & \text{if }c_0 = 0;
    \end{cases}\]
  \[\frac{\partial L}{\partial \pmfLike_j}(\pmfLike, \pmfInfty, \mu, \lambda) = \frac{c_j}{\pmfLike_j} - \mu - \lambda v_j \quad\text{ for } j \in [m];\]
  \[\frac{\partial L}{\partial \pmfInfty}(\pmfLike, \pmfInfty, \mu, \lambda) =
    \begin{cases}
      \frac{c_{m + 1}}{\pmfInfty} - \lambda & \text{if }c_{m+1} > 0,\\
      - \lambda & \text{if } c_{m+1} = 0.
    \end{cases}\]
  Solving the KKT conditions yields:
  \begin{enumerate}
  \item If $c_{m + 1} = 0$ and ${\sum_{j= 1}^mv_jc_j} \le {\sum_{j = 0}^mc_j}$, then
    \begin{align*}
      \pmfLike_j = \frac{c_j}{\mu}\quad\text{for } 0\le j\le m;
    \end{align*}
    \begin{align*}
      \pmfInfty = 1 - \frac{\sum_{j = 1}^mv_jc_j}{\mu};\quad \mu = n;\quad \lambda = 0.
    \end{align*}
  \item Otherwise, if $c_0 = 0$ and ${\sum_{j = 1}^mc_j/v_j} \le {\sum_{j = 1}^{m + 1}c_j}$, then
    \begin{align*}
      \pmfLike_j = \frac{c_j}{\lambda v_j}  \quad\text{for } 1\le j\le m;
    \end{align*}
    \begin{align*}
      \pmfLike_0 = 1 - \frac{\sum_{j = 1}^mc_j/v_j}{\lambda};
    \end{align*}
    \begin{align*}
      \pmfInfty = 0;\quad
      \mu = 0; \quad\lambda = n.
    \end{align*}
  \item Otherwise, $\mu > 0$, $\lambda > 0$ are determined by
    solving
    \begin{equation}
      \label{eq:c1_sum}
      \sum_{j = 0}^m\frac{c_j}{\mu + \lambda v_j} = 1,
    \end{equation}
    \begin{equation}
      \label{eq:c2sum}
      \sum_{j = 1}^{m}\frac{c_jv_j}{\mu + \lambda v_j} + \frac{c_{m + 1}}{\lambda} = 1,
    \end{equation}
    and for $0\le j\le m$,
    \[\pmfLike_j = \frac{c_j}{\mu + \lambda v_j},\quad b = \frac{c_{m + 1}}{\lambda}.\]
  \end{enumerate}
  Multiplying both sides of \eqref{eq:c1_sum} by $\mu$ and both sides of \eqref{eq:c2sum} by $\lambda$ and adding the respective sides of the two equations obtained, yields $\mu + \lambda = \sum_{j=0}^{m+1} c_j = n$.  The above conditions can be expressed in terms of the variables $R_i$, and then replacing $\mu$ by $n(1-\lambda_n)$ and $\lambda$ by $n \lambda_n$ yields the proposition.
\end{IEEEproof}

\begin{IEEEproof}[Proof of Corollary~\ref{cor:ML_estimator_alt}]
  Corollary~\ref{cor:ML_estimator_alt} is deduced from Proposition~\ref{prop:ML_estimator} as follows.  If $R_i = 1$ for $1\leq i \leq n$ then the corollary gives that both $\hat{F_0}$ and $\hat{F}_1$ have all their mass at $r = 1$, in agreement with Proposition~\ref{prop:ML_estimator}.  So for the remainder of the proof suppose $R_i\neq 1$ for some $i$.

  Consider the three cases of Proposition~\ref{prop:ML_estimator}.  If case 1) holds then $\phi_n(0)=1$ and $\phi'_n(0) =\frac 1 n \sum_{i=1}^n (1-R_i) \geq 0$.  Also, $R_i < \infty$ for $1\leq i \leq n$.  Since $R_i\not\in \{1, \infty\}$ for at least one value of $i$, $\phi_n$ is strictly convex over $[0, 1]$.  Therefore, $\phi_n(\lambda) > 1$ for $\lambda \in (0, 1]$.  Thus, $\lambda_n$ defined in the corollary is given by $\lambda_n=0$, and the corollary agrees with Proposition~\ref{prop:ML_estimator}.

  If case 2) holds then $\phi_n(1)\leq 1$.  Thus, $\lambda_n$ defined in the corollary is given by $\lambda_n = 1$, and the corollary agrees with Proposition~\ref{prop:ML_estimator}.

  If neither case 1) nor case 2) holds, then $\lambda_n$ in the corollary is the same as $\lambda_n$ in Proposition~\ref{prop:ML_estimator}, and the corollary again agrees with Proposition~\ref{prop:ML_estimator}.
\end{IEEEproof}

\subsection{From Pointwise to Uniform Convergence of CDFs}
\label{app:CDF_pointwise_vs_unif}

The following basic lemma shows that uniform convergence of a sequence $\ind{F_n}{n\geq 1}$ of CDFs to a fixed limit is equivalent to pointwise convergence of both the sequence and the corresponding sequence of left limit functions, at each of a suitable countably infinite set of points.  The CDFs in this section may correspond to probability distributions with positive mass at $-\infty$ and/or $\infty$.
\begin{lemma}[Finite net lemma for CDFs]
  \label{lemma:net}
  Given a CDF $F$ and any integer $L\geq 1$, there exist $c_1, \dots , c_{L-1} \in \reals\cup \{-\infty, \infty\}$ such that for any CDF $G$, $d_{KS}(F,G) \leq \delta + \frac 1 L$ where
  \begin{align*}
    \delta = \max_{1\leq \ell\leq L-1} \max \{|F(c_\ell)-G(c_\ell)|, |F(c_\ell-)-G(c_\ell-)|\}.
  \end{align*}
\end{lemma}
\begin{IEEEproof}
  Let $c_\ell = \min\set{c \in \reals \cup \{-\infty, \infty\}}{F(c) \geq \frac \ell L}$ for $1\leq \ell \leq L-1$.  Also, let $c_0=-\infty$ and $c_L=\infty$.  The fact $F(c_{\ell+1}-)-F(c_\ell) \leq \frac 1 L$ for $0\leq \ell \leq L-1$ and the monotonicity of $F$ and $G$ implies the following.  For $0\leq \ell \leq L-1$ and $c \in (c_\ell, c_{\ell+1})$,
  \begin{align*}
    G(c) \geq G(c_\ell) \geq F(c_\ell) -\delta \geq F(c) -\delta - \frac 1 L
  \end{align*}
  and similarly
  \begin{align*}
    G(c) \leq G(c_{\ell+1}-) \leq F(c_{\ell+1}-) + \delta \leq F(c) + \delta + \frac 1 L.
  \end{align*}
  Since $\reals \subset \{c_1, \dots , c_{L-1}\} \cup \left( \cup_{\ell=1}^{L-1} (c_\ell, c_{\ell+1})\right)$, it follows that $|F(c)-G(c)| \leq \delta + \frac 1 L$ for all $c\in\reals$, as was to be proved.
\end{IEEEproof}

\begin{cor}
  \label{cor:CDF_pointwise_vs_unif}
  If $F$ is a CDF, there is a countable sequence $\ind{c_\ell}{\ell \geq 1}$ such that, for any sequence of CDFs $\ind{F_n}{n\geq 1}$, $d_{KS}(F,F_n)\to 0$ if and only if $F_n(c_\ell)\to F(c_\ell)$ and $F_n(c_\ell-)\to F(c_\ell-)$ as $n\to \infty$ for all $\ell\geq 1$.
\end{cor}
\begin{IEEEproof}
  Given $F$, let $\ind{L_j}{j\geq 1}$ be a sequence of integers converging to $\infty$.  For each $j$, Lemma~\ref{lemma:net} implies the existence of $L_j-1$ values $c_\ell$ with a specified property.  Let the infinite sequence $\ind{c_\ell}{\ell \geq 1}$ be obtained by concatenating those finite sequences.
\end{IEEEproof}

\subsection{Proof of Consistency of ML Estimator}
\label{app:consistency}
The proof of Proposition~\ref{prop:M_properties} will be given using a series of lemmas.
\begin{lemma}
  \label{lemma:key_phi_fact}
  Let $\phi$ be defined by \eqref{eq:phi_def} for the CDF $F$ of a probability measure supported over $[0,\infty].$ Then $\phi$ is a continuous and convex function over $[0,1]$ and $\phi(\lambda)\leq \frac 1 {1-\lambda}$ for $0\leq \lambda < 1.$ ($\phi(1)=\infty$ is possible). If $F(\{1\} )< 1$ and $\phi(0) = 1$ then $\phi$ is strictly convex over $[0,1].$
\end{lemma}
\begin{IEEEproof}
  Let $g(\lambda,r) = \frac 1 {1- \lambda + \lambda r},$ so $\phi(\lambda) = \int_0^\infty g(\lambda, r) dF(r).$ Since $g(\lambda,r)\leq \frac 1 {1-\lambda}$ for all $r\geq 0$ and $0\leq \lambda < 1$ it follows that $\phi(\lambda)\leq \frac 1 {1-\lambda}.$   The function $g(\lambda,r)$ is bounded and continuous in $\lambda$ for $\lambda \in [0,1-\epsilon]$ for any $\epsilon > 0,$  so by the bounded convergence theorem, $\phi$ is continuous over the set $[0,1).$  Similarly, the function $\lambda \mapsto \int_1^\infty g(\lambda,r) dF(r)$ is a bounded continuous function over $\lambda \in [0,1].$  The function $g(\lambda,r)$ is monotone increasing in $\lambda$ for $r\in [0,1]$ so by the monotone convergence theorem, the function
   $\lambda \mapsto \int_0^1 g(\lambda,r) dF(r)$ is continuous at $\lambda = 1.$  Therefore $\phi$ is also continuous at $\lambda =1,$ and is hence continous over $[0,1]$ as claimed.

  Note that $\phi(0) = \int_0^{\infty} 1~ dF(r) = 1 - F(\{\infty\}).$ If $\phi(0) = 1$, then $F(\{\infty\})=0$ and if also $F(\{1\}) < 1$ then $F([0,1)\cup (1,\infty))>0$ so $g(\lambda,r)$ is strictly convex in $\lambda$ for $r$ in a set of strictly positive probability under $F$, so $\phi$ is strictly convex under those conditions.
\end{IEEEproof}

\begin{lemma}
  \label{lemma:cases_phi}
  (a) If
  \begin{align}
    \int_{[0,\infty]}  r dF(r) \leq 1  \label{eq:r_condition}
  \end{align}
  then $F=F_0$ and if also $F_0\neq F_1$ then $\beta=0$ and $\phi(\lambda) >1$  for $0 <\lambda \leq 1.$ \\
  (b) If
  \begin{align}
    \int_{[0,\infty]} \frac 1 r dF(r) \leq 1  \label{eq:1/r_condition}
  \end{align}
  then $F=F_1$ and $\beta=1.$  Moreover, if also $F_0  \neq F_1$ then $\phi(\lambda) < 1$ for $0< \lambda < 1.$ \\
  (c) If neither \eqref{eq:r_condition} nor \eqref{eq:1/r_condition} hold then $0<\beta< 1.$  Moreover, $\phi(\lambda) < 1$ for $0< \lambda < \beta$ and $\phi(\lambda) > 1$ for $\beta< \lambda < 1.$
\end{lemma}

\begin{IEEEproof}
  Proof of (a): Suppose \eqref{eq:r_condition} holds.  It implies that $F(\{\infty\})=0$ so $\phi(0)=1$ and also $\phi'(0) = 1 - \int_0^\infty r dF(r) \geq 0.$ Furthermore, if $F_0 \neq F_1$ then $\phi$ is strictly convex by Lemma \eqref{eq:r_condition} so $\phi(\lambda) > 1$ for $\lambda \in (0,1]$ and $\beta=0,$ so $F=F_0$ by \eqref{eq:F0_fromF}.  If $F_0=F_1$ then $F=F_0=F_1.$ In either case, $F=F_0.$

  Proof of (b): Suppose \eqref{eq:1/r_condition} holds.  Then $\phi(1) =\int_0^{\infty} \frac 1 r F(dr)\leq 1$ so $\beta=1.$ So $F=F_1$ by \eqref{eq:F1_fromF}. The last statement of (b) follows from Lemma~\ref{lemma:key_phi_fact}.

  Proof of (c): Suppose neither \eqref{eq:r_condition} nor \eqref{eq:1/r_condition} holds.  Note that $\phi(0)= \int_0^\infty dF(r) = 1 - F(\{\infty\}).$ So either $\phi(0)< 1$ or ($\phi(1)=1$ and
  \begin{align*}
    \phi'(0)= 1 - \int_0^{\infty} r F(dr) = 1 - \int_{[0,\infty]} r F(dr) < 0).
  \end{align*}
  Either way, $\phi(\lambda)<0$ for sufficiently small positive values of $\lambda$, $\phi$ is convex by Lemma~\ref{lemma:key_phi_fact}, and $\phi(1)= \int_0^{\infty} \frac 1 r F(dr) > 1.$ Therefore there is a unique value of $\lambda \in (0,1)$ such that $\phi(\lambda)=1,$ and that must equal $\beta.$ The final statement also follows.
\end{IEEEproof}

We here begin the proof of the continuity assertion in Proposition~\ref{prop:M_properties}.  So let $F$ and $F_n$ for $n\geq 1$ be CDFs for probability distributions supported on $[0,\infty]$ such that $d_{KS}(F,F_n) \rightarrow 0.$ Let $(F_0,F_1,\ROC) = {\cal M}(F)$ and let $\phi$ and $\beta$ also correspond to $F$ as in the definition of ${\cal M}(F).$ Similarly, for each $n\geq 1,$ let $(F_{0,n},F_{1,n},\ROC_n) = {\cal M}(F_n)$ and let $\phi_n$ and $\beta_n$ also correspond to $F_n$ as in the definition of ${\cal M}(F_n).$ It is sufficient to show that $d_{KS}(F_k, F_{k,n}) \rightarrow 0$ for $k\in\{0,1\},$ because, by Lemma~\ref{lemma:BHT_metric_bound}, this implies that $L(\ROC,\ROC_n)\rightarrow 0.$ By the finite net lemmma for CDFs, Lemma~\ref{lemma:net}, it suffices to prove pointwise convergence of CDFs and their left limits -- i.e. for any fixed $\tau > 0$ that $F_{k,n}(\tau) \rightarrow F_k(\tau)$ and $F_{k,n}(\tau-) \rightarrow F_k(\tau-)$ for $k=0,1.$

The following lemma is a special case of the product formula in semimartingale stochastic calculus, which for two right-continuous-with-left-limits functions $X$ and $Y$ with bounded variation states (\cite{WongHajek85}, Section~6.6): $X_tY_t=X_0Y_0+\int_0^t X_{s-}dY(s) + \int_0^t Y_{s-}dX(s) + \sum_{0 < s \leq t} \Delta X_s \Delta Y_s.$  If one of the functions is continuously differentiable (as in the following lemma) then $X_tY_t=X_0Y_0+\int_0^t X_{s}dY(s) + \int_0^t Y_{s}dX(s).$

\begin{lemma}(Integration by parts)
  Let $h$ be a continuously differentiable function on $[0,\infty)$ and let $F$ be a CDF for a probabilty measure on $[0,\infty].$ Then for any closed interval $[a,b] \subset [0,\infty),$
  \begin{align*}
    \int_a ^ b  h(r) dF(r) = F(b)h(b) - F(a-)h(a) - \int_a^b  h'(\tau) F(\tau) d\tau
  \end{align*}
\end{lemma}

\begin{lemma}  \label{lemma:phi_converge}
  For $0 \leq \lambda < 1$
  \begin{align}
    \left| \phi(\lambda)-\phi_n(\lambda)\right| \leq \frac 1 {1-\lambda}d_{KS}(F_n,F)
  \end{align}
  Thus, $\phi_n$ converges to $\phi$ uniformly on intervals of the form $[0,\delta]$ for any $\delta$ with $0 < \delta < 1.$
\end{lemma}
\begin{IEEEproof}
  By continuity at $\lambda=0$ it suffices to prove the lemma for $0 < \lambda < 1.$ So fix $\lambda$ with $0 < \lambda < 1$ and define $h(r)=\frac 1 {1-\lambda +\lambda r}.$ Then by integration by parts over $[0,b]$ and taking the limit $b\to \infty,$ and using the facts $F(0-)=\lim_{b\to\infty} h(b)= 0$ and $h'(r) < 0,$
  \begin{align*}
    \phi(\lambda) = - \int_0^{\infty} h'(r) F(r) dr,
  \end{align*}
  and $\phi_n$ is determined by $F_n$ in the same way.  Thus
  \begin{align*}
    |\phi(\lambda) - \phi_n(\lambda)|& \leq  -\int_0^{\infty} h'(r) |F(r)-F_n(r)| dr \\
                                     & \leq \left(-\int_0^{\infty} h'(r) dr\right)d_{KS}(F_n,F)\\
                                     &=h(0)d_{KS}(F_n,F)
  \end{align*}
  which yields the lemma.
\end{IEEEproof}

\begin{lemma}  \label{lemma:beta_convergence}
  If $F_0 \neq F_1$ then $\beta_n \rightarrow \beta.$
\end{lemma}
\begin{IEEEproof}
  Suppose $F_0\neq F_1$ and consider the three cases defined in Lemma~\ref{lemma:cases_phi}. In case (a), $\phi(r_o)>1$ for any $r_o>0.$ It follows that $\phi_n(r_o)>1$ for all sufficiently large $n.$ Since $\phi_n$ is a convex function with $\phi_n(0)\leq 1$ and $\phi_n(r_o) > 1$ it must be that $\phi_n(r)> 1$ for $r>r_o.$ Thus, $\beta_n < r_o$ for all sufficiently large $n$.  Since $r_o$ was arbitrary, $\beta_n\rightarrow 0=\beta.$

  In case (b), $\phi(r_1)<1$ for any $r_1 \in (0,1).$ It follows that $\phi_n(r_1)<1$ for all sufficiently large $n.$ Thus, $\beta_n \geq r_1$ for all sufficiently large $n$.  Since $r_1$ was arbitrary, $\beta_n\rightarrow 1=\beta.$

  In case (c), $0<\beta < 1.$ If $0 < \epsilon < \min\{\beta,1-\beta\}$ then $\phi(\beta-\epsilon) < 1 < \phi(\beta+\epsilon).$ Therefore, for all sufficiently large $n,$ $\phi_n(\beta-\epsilon) < 1 < \phi_n(\beta+\epsilon),$ which implies $|\beta-\beta_n| < \epsilon$ for sufficiently large $n.$ Since $\epsilon$ was arbitrary, $\beta_n\rightarrow 0=\beta.$
\end{IEEEproof}

\begin{IEEEproof}[Completion of proof of Proposition~\ref{prop:M_properties}]
  Let $(F_0,F_1,\ROC)$ be a BHT and $\alpha \in [0,1]$ and let $F=(1-\alpha) F_0 + \alpha F.$ Equality ${\cal M}(\hat{F}) = (\hat{F}_{0,ML}, \hat{F}_{1,ML}, \ROCML )$ follows from comparing the definition of $\cal M(\hat{F})$ to the description of $\ROCML$ in Lemma~\ref{cor:ML_estimator_alt}. (For that it should be noted that the terms in $\hat{F}_0^c(\tau)$ with $R_i=\infty$ are zero because if $R_i=\infty$ for some $i$ then $\lambda_n >0.$) Next it will be shown that ${\cal M}(F)=(F_0,F_1,\ROC).$ The result is easily verified if $F(\{1\})=1$ or equivalently if $F_0=F_1$ so assume for the remainder of the proof that $F_0 \neq F_1.$ Since \eqref{eq:F0_fromF} and \eqref{eq:F1_fromF} reduce to \eqref{eq:alphaF0_fromF} and \eqref{eq:alphaF1_fromF}, respectively, if $\beta=\alpha,$ it suffices to prove $\beta = \alpha,$ where $\beta$ appears in the definition of ${\cal M}(F).$

  If $\alpha=0$ then $F=F_0$ and $rdF_0(r) = dF_1(r)$ for $0 \leq r < \infty$ so that \eqref{eq:r_condition} holds. Lemma~\ref{lemma:cases_phi} implies $\beta=0=\alpha.$ If $\alpha=1$ then $F=F_1$ and $dF_0(r)=\frac 1 r dF_1(r)$ so that \eqref{eq:1/r_condition} holds.  Lemma~\ref{lemma:cases_phi} implies $\beta=1=\alpha.$ If neither \eqref{eq:r_condition} nor \eqref{eq:1/r_condition} hold then by Lemma~\ref{lemma:cases_phi} $\beta$ is the unique value with $0 < \beta < 1$ such that $\phi(\beta)=1.$ Since
  \begin{align*}
    \phi(\alpha)=\int_0^{\infty} \frac 1 {1-\alpha+ \alpha r}(1-\alpha+ \alpha r)dF_0(r) =1
  \end{align*}
  it must again be that $\alpha=\beta.$ Thus, if $F_0\neq F_1$ then $\beta=\alpha.$ The proof of Proposition~\ref{prop:M_properties}(i) is complete.

  Turn to the proof of Proposition~\ref{prop:M_properties}(ii).  Using the triangle inequality we have for any $\tau > 0,$
  \begin{align*}
    | F_0^c(\tau)-F_{0,n}^c(\tau)| & = \left| \int_{\tau+}^\infty \frac 1 {1-\beta+\beta r } dF(r) -\right. \\
    &\left. ~~~~\int_{\tau+}^\infty \frac 1 {1-\beta_n+\beta_n r } dF_n(r)   \right| \\
    &  \leq \delta_{1,n} + \delta_{2,n}
  \end{align*}
  where
  \begin{align*}
    \delta_{1,n}& = \max_{r \in [\tau,\infty)} \left| \frac 1 {1- \beta + \beta r} - \frac 1 {1- \beta_n + \beta_n r}\right| \rightarrow 0 \\
    \delta_{2,n}& = \left| \int_{\tau+}^\infty \frac 1 {1- \beta + \beta r} dF_n(r) - \int_{\tau+}^\infty \frac 1 {1- \beta + \beta r} dF(r) \right| \\
                & \leq \frac 1 {1-\beta + \beta \tau} d_{KS}(F_n,F) \rightarrow 0,
  \end{align*}
  where we used the fact $\beta_n\rightarrow \beta$ to imply $\delta_{1,n}\rightarrow 0$ and for the last inequality we applied integration by parts with $h(r)=\frac 1 {1-\beta + \beta r}$ as in the proof of Lemma~\ref{lemma:phi_converge}.  Thus $F_{0,n}(\tau) \rightarrow F_0(\tau).$ The proofs that $F_{1,n}(\tau) \rightarrow F_1(\tau)$ and $F_{k,n}(\tau-) \rightarrow F_k(\tau-)$ for $k\in \{0,1\}$ are similar and omitted.  That last assertion of Proposition~\ref{prop:M_properties}(ii) follows from Lemma~\ref{lemma:beta_convergence} that $\beta_n\to \beta$ together with the fact $\beta=\alpha$ as proved above.  The proof of Proposition~\ref{prop:M_properties} is complete.
\end{IEEEproof}

\begin{exam}
  \label{exam:not_Lip}
  While the mapping $\cal M$ is continuous it is not Lipschitz continuous as indicated in this example.  Let $\epsilon$ be a parameter with $0 \leq \epsilon < \frac 1 4.$ The probability distributions $F^{\epsilon}$, $F_0^{\epsilon}$, and $F_1^{\epsilon}$ in this example are each supported on the set $\{0,2,\infty\}$ with the probabilities assigned to the three possible values given as follows:
  \begin{align*}
    F^{\epsilon} \leftrightarrow &  \left(\frac 1 2 + \epsilon, \frac 1 2 - 2\epsilon, \epsilon\right) \\
    F^{\epsilon}_0 \leftrightarrow & \left(\frac{ \frac 1 2 + \epsilon}{1-\alpha_{\epsilon}},
                                     \frac{ \frac 1 2 - 2\epsilon}{1+\alpha_{\epsilon}}, 0\right)  \\
    F_1^{\epsilon} \leftrightarrow & \left(0,
                                     \frac{ 1 - 4\epsilon}{1+\alpha_{\epsilon}}, \frac{\epsilon}{\alpha_{\epsilon}} \right)
  \end{align*}
  where $\alpha_{\epsilon} = \frac{\sqrt{9\epsilon^2 + 4\epsilon} -3\epsilon} 2.$ It can be checked that for each $\epsilon$, ${\cal M}(F^{\epsilon}) = (F_0^{\epsilon}, F_1^{\epsilon},\ROC^{\epsilon})$, where $\ROC^{\epsilon}$ is the ROC curve associated with $F_0^{\epsilon}$ or, equivalently, $F_1^{\epsilon}.$ Specifically, $\ROC^{\epsilon}$ has three linear segments: a vertical segment going up from $(0,0)$ to $(0,F_1^{\epsilon}(\{\infty\})),$ a segment with slope 2 rising to height one, and a horizontal segment.  Note that $\alpha_{\epsilon} \asymp \sqrt{\epsilon}$ as $\epsilon \to 0.$ Furthermore $d_{KS}(F,F^{\epsilon}) = \epsilon$ and $L(\ROC_0, \ROC_{\epsilon}) = \frac {\epsilon}{3\alpha_{\epsilon}} \asymp \frac{\sqrt{\epsilon}} 3.$ Thus, the ratio $L(\ROC_0, \ROC_{\epsilon}) / d_{KS}(F,F^{\epsilon})$ is unbounded as $\epsilon \to 0.$ Similarly, $d_{KS}(F_1,F_1^{\epsilon})/d_{KS}(F,F^{\epsilon})$ is unbounded.  This example is centered on a situation that most of the observations are generated under the same hypothesis, namely, $H_0.$
\end{exam}

\subsection{Derivation of Expressions for $\AUC$ and $\AUCML$}
\begin{IEEEproof}[Proof of Proposition~\ref{prop:AUC_formulas}]
  (Proof of 1) Let $R_1 \leq \dots \leq R_n$ denote the ordered observed likelihood ratio samples. Then the region under $\ROCML$ can be partitioned into a union of trapezoidal regions, such that there is one trapezoid for each $R_i$ such that $R_i<\infty.$ The trapezoids are numbered from right to left.  If a value $v_j \in (0,\infty)$ is taken on by $c_j$ of the samples, then the union of the trapezoidal regions corresponding to those samples is also a trapezoidal region.

  The area of the $i$th trapezoidal region is the width of the base times the average of the lengths of the two sides.  The width of the base is $\frac 1 n \cdot \frac{1}{1 - \lambda_n + \lambda_n R_i}$, corresponding to a term in $\hat{F_0}$.  The length of the left side is $\frac 1 n \cdot\sum_{i':i'>i} \frac{1}{1 - \lambda_n + \lambda_n R_{i'}}$, and the length of the right side is greater than the length of the left side by $\frac 1 n \cdot \frac{1}{1 - \lambda_n + \lambda_n R_i}$.  Summing the areas of the trapezoids yields:
  \begin{align*}
    & \AUCML = \frac{1}{n^2} \sum_{i=1}^{n} \bigg\{ \frac 1 { 1- \lambda_n + \lambda_n R_i} \\
    & \cdot \left( \left( \sum_{i'=i+1}^{n} \frac{ R_{i'} }{1- \lambda_n +\lambda_n R_{i'})} \right) + \frac 1 2 \frac{ R_i }{1 -\lambda_n + \lambda_n R_i)} \right) \bigg\},
  \end{align*}
  which is equivalent to the expression given in 1) of the proposition.

  (Proof of 2) The consistency of $\AUCML$ follows from Corollary~\ref{cor:consistency}, the consistency of $\ROCML$.

  (Proof of 3) Let $\tau(p)$ and $\eta(p)$ denote values $\tau(p) \in [0, \infty)$ and $\eta(p)\in [0, 1]$ such that $F_0^c(\tau(p), \eta(p))=p$.  Then
  \begin{align*}
    \AUC & = \int_0^1 \ROC(p) \d p =  \int_0^1  F_1^c(\tau(p),\eta(p)) \d p  \\
         & = \int_0^1  \left( \eta(p)F_1^c(\tau(p))+ (1-\eta(p))F_1^c(\tau(p)-) \right)\d p\\
         &\stackrel{(a)}{=}  \int_0^1 \frac{F_1^c(\tau(p))+ F_1^c(\tau(p)-)} 2 \d p\\
         &\stackrel{(b)}{=}  \E_0\left[\frac{F_1^c(R) + F_1^c(R-)} 2 \right] \\
         & = \E_0 \left\{ \frac{  \int_{R+}^\infty  r' \d F_0(r') +
           \int_{R}^\infty  r' \d F_0(r')} 2 + F_1(\{\infty\})  \right\}  \\
         & = \E_0\left[ R' \left(\indicator{\{R'>R\}} + \frac 1 2 \indicator{\{R'=R\}}\right)\right] + F_1(\{\infty\}) \\
         & = \frac 1 2 \E_0[\max\{R,R'\}] + F_1(\{\infty\})  \\
         & = \frac 1 2 \E_0[\max\{R,R'\}] + 1-\E_0[R]  \\
         & = 1 - \frac 1 2 \E_0[R + R'  - \max\{R,R'\}] \\
         & = 1 - \frac 1 2 \E_0[\min\{R,R'\}],
  \end{align*}
  where (a) follows from the fact that $\ROC(p)$ is affine over the maximal intervals of $p$ such that $\tau(p)$ is constant, so the integral is the same if $\ROC(p)$ is replaced over each such interval by its average over the interval, and (b) follows from the fact that if $U$ is a random variable uniformly distributed on the interval $(0, 1)$, then the CDF of $\tau(U)$ is $F_0$ because for any $c \geq 0$, $\Pr\{ \tau(U) > c \} = \Pr\{ U \leq F^c_0(c)\} = F^c_0(c)$.  This establishes \eqref{eq:AUC_formula_1} and \eqref{eq:AUC_formula_2}.

  (Proof of 4) This follows from \eqref{eq:AUC_formula_1} and the fact the CDF of $R$ and $R'$ satisfies $\d F(r)=(1 - \alpha + \alpha r) \d F_0(r)$ over $[0, \infty)$ and $F(\{\infty\})=\alpha F_1(\{\infty\})$.
\end{IEEEproof}

\section{Proofs for Section~\ref{sec:split_estimator} -- the split and fused estimators}

\subsection{Legendre transforms}
\label{sec:Legendretransforms}

This section provides background for the proof of Proposition~\ref{prop:finite_sample_bnds} in the next section.   We shall work with the Legendre transforms of ROCs and the pseudo ROCs defined in Section~\ref{sec:split_estimator}. Legendre transforms are usually defined for convex functions.  For concave functions we use a variation of the usual Legendre transform.  A {\em proper} concave function on $\reals$ is a concave function with values in $\reals \cup \{ -\infty \}$ (i.e. in $[-\infty, \infty)$) that is not identically $-\infty$ and is upper semicontinuous.  Similarly, a proper convex function is the negative of a proper concave function.  Given a proper concave function $f$, we define its Legendre transform by
\begin{align*}
  f^*(r) = \sup_{p\in \reals}  f(p) - pr ~~\mbox{for } r\in \reals
\end{align*}
A geometric interpretation is that $f^*(r)$ is the $y$-axis intercept of the line of slope $r$ tangent to the graph of $f.$ If LT denotes the usual Legendre transform of proper convex functions defined by $LT(g)(r) = \sup_x xr- g(x)$, then $f^*$ here can be expressed as $f^*(p)=LT(-f)(-p).$ Some key properties of the Legendre transform are collected into the following lemma, stated without proof. The last item in the lemma follows readily from the property listed just before it.

\begin{lemma}[Properties of Legendre transform of proper concave functions]
  \begin{enumerate}
  \item (Inversion) If $f$ is a proper concave function, then $f^*$ is a proper convex function and $f(p) = \inf_{r \in \reals} f^*(r) + pr.$ This is a version of the well known fact that a proper concave function is the pointwise infimum of the collection of all affine functions that dominate it.

  \item (Inversion for monotone $f$) If $f$ is a proper concave function and nondecreasing, then $f^*(r)=+\infty$ for $r < 0$, so that $f(p) = \inf_{r \geq 0} f^*(r) + pr.$

  \item (Order preserving) If $f$ and $g$ are proper concave functions then $f\geq g$ (pointwise) if and only if $f^* \geq g^*$ (pointwise). (With the convention that
    $-\infty \geq -\infty$ and $\infty \geq \infty.$)

  \item (Isometry in sup norm) If $f$ and $g$ are proper concave functions, $\|f - g\|_{\infty} = \|f^*-g^*\|_{\infty}.$ (With the convention that $-\infty - (-\infty)=0$ and $\infty-\infty=0.$)

  \item (Transform under shifts) If $f$ is a proper concave function, then the transform of $x\mapsto f(x-\epsilon) + \epsilon$ is $r\mapsto f^*(r) + \epsilon (1+r).$

  \item (\levy\ distance) If $f$ and $g$ are nondecreasing, proper concave functions, then the \levy\ distance between them is given by
    \begin{align} \label{eq:L_for_transforms}
      L(f, g) = \sup_{r \geq 0}  \frac{|f^*(r)-g^*(r)|}{1+r}.
    \end{align}
  \end{enumerate}
\end{lemma}

\subsection{Proof of Performance Bound for Split and Fused Estimators}

Proposition~\ref{prop:finite_sample_bnds} is proved in this section.  The domain of the mappings ${\cal R}_{UR}$ and ${\cal R}_{LL}$ and their clean versions ${\cal R}_{URC}$ and ${\cal R}_{LLC}$ can be extended to the family of all CDFs $F$ supported by $[0,\infty],$ under the following restriction:
\begin{assump}
  \label{assump:F_lambda_exclusion}
  If $\lambda =0$ then $F(\{\infty\})=0$ and if $\lambda=1$ then $F(0)=0.$
\end{assump}
Note that Assumption~\ref{assump:F_lambda_exclusion} is satisfied by the pairs $(\hat{F} , \lambda_n)$ arising in the ML estimator.

The extensions are described by specifying the Legendre transforms of the ROC curves.  Appendix~\ref{sec:Legendretransforms} describes the properties of Legendre transforms we shall use.  Using the interpretation that the value of the transform at a value $r\geq 0$ is the value of the $y$-intercept for the line of slope $r$ tangent to the curve, the following expressions for the Legendre transforms of ${\cal R}_{UR}(\hat{F},\lambda)$ and ${\cal R}_{LL}(\hat{F},\lambda)$ are readily obtained.  For $r\geq 0$
\begin{align}
  {\cal R}_{UR}^*(\hat{F},\lambda)(r) & = 1 - r + \frac 1 n \sum_{j=1}^n \frac{(r - R_j)_+}{1-\lambda +\lambda R_j} \label{eq:LTURn}\\
  {\cal R}_{LL}^*(\hat{F},\lambda)(r) & = \frac 1 n \sum_{j=1}^n \frac{(R_j - r)_+}{1-\lambda +\lambda R_j}.
                                        \label{eq:LTLLn}
\end{align}

The mappings ${\cal R}_{UR}(\hat{F},\lambda)$ and ${\cal R}_{LL}(\hat{F},\lambda)$ can be extended to be defined for $F$ being the CDF of any probability distribution supported by $[0,\infty]$ and $\lambda \in [0,1]$ subject to Assumption~\ref{assump:F_lambda_exclusion} by the following definitions for their Legendre transforms:
\begin{align*}
  {\cal R}^*_{UR}(F,\lambda)(r) &= 1 -r + \int_0^r \frac{r-s}{1-\lambda + \lambda s} dF(s) \mbox{ for } r\geq 0.  \\
  {\cal R}^*_{LL}(F,\lambda)(r) &= \int_r^\infty \frac{s-r}{1-\lambda + \lambda s} dF(s) + \frac{F(\{\infty\})}{\lambda} \mbox{ for } r\geq 0.
\end{align*}
Define the associated clean versions of ${\cal R}_{UR}$ and ${\cal R}_{LL}$ by ${\cal R}_{URC}(F,\lambda) = T^{conc}\circ T^{proj} \left( {\cal R}_{UR}(F,\lambda)\right)$ and ${\cal R}_{LLC}(F,\lambda) = T^{conc}\circ T^{proj} \left( {\cal R}_{LL}(F,\lambda)\right).$

\begin{lemma}
  \label{lemma:IBP}
  Let $F$ and $G$ be CDFs on $[0,\infty]$ and let $C$ be a nonincreasing, nonnegative, right continuous function on $[0,\infty).$ Then
  \begin{align*}
    \sup_{r\geq 0} \int_0^r C(s)(dF(s) - dG(s)) \leq C(0) d_{KS}(F,G).
  \end{align*}
\end{lemma}
\begin{IEEEproof}
  By integration by parts, for any $r\geq 0,$
  \begin{align}
    \label{eq:IBP}
    \int_0^r C(s)(dF(s) - dG(s))& = C(0) \left[ \int_0^r (F(s)-G(s)) \frac{-dC(s)}{C(0)} + (F(r)-G(r))\frac{C(r)}{C(0)} \right]
  \end{align}
  The quantity in square brackets on the righthand side of \eqref{eq:IBP} is a weighted average of $F(s)-G(s)$ over $[0,r]$ (with total weight one) so the bound in the Lemma follows.
\end{IEEEproof}

\begin{lemma}
  \label{lemma:est_contraction}
  (a) For $\lambda\in [0,1)$ fixed, the mapping $F\mapsto {\cal R}_{URC}(F,\lambda)$ is a $\frac 1 {(1-\lambda)}$-Lipschitz continuous mapping from the space of CDFs with the $d_{KS}$ metric to the space of ROC curves with the \levy\ metric.
  (b) For $\lambda\in (0,1]$ fixed, the mapping $F\mapsto {\cal R}_{LL}(F,\lambda)$ is a $\frac 1 \lambda$-Lipschitz continuous mapping from the space of CDFs with $d_{KS}$ metric to the space of ROC curves with the \levy\ metric.
\end{lemma}

\begin{IEEEproof}
  Both $T^{proj}$ and $T^{conc}$ are contractions in the \levy\ metric (the contractive property of $T^{conc}$ is part of Proposition~\ref{prop:dkw-empirical}).  Thus, it suffices to prove the Lipschitz property for the mappings ${\cal R}_{UR}(F,\lambda)$ and ${\cal R}_{UR}(F,\lambda).$ We have
\begin{align*}
  L({\cal R}_{UR}(F,\lambda), {\cal R}_{UR}(G,\lambda)) & \stackrel{(a)}{=} \sup_{r\geq 0} \frac{\left|{\cal R}^*_{UR}(F,\lambda)(r) - {\cal R}_{UR}^*(G,\lambda)(r)\right|}{1+r} \\
                                                        & \stackrel{(b)}{=} \sup_{r\geq 0} \bigg| \int_0^r  \frac{(r-s)(dF(s)-dG(s))}{(1+r)(1-\lambda + \lambda s)}  \bigg|  \\
                                                        & \stackrel{(c)}{\leq}  \frac{d_{KS}(F,G)}{1-\lambda},
\end{align*}
where (a) follows by the formula \eqref{eq:L_for_transforms} for \levy\ distance in terms of the transforms, (b) follows from the definitions of the two Legendre transforms, and (c) follows from Lemma~\ref{lemma:IBP}.  The proof of Lemma~\ref{lemma:est_contraction}(a) is complete and the proof of Lemma~\ref{lemma:est_contraction}(b) follows from (a) by symmetry: swapping $H_0$ and $H_1$, $\lambda$ and $1-\lambda,$ and $r$ and $1/r$ maps the problem to itself.
\end{IEEEproof}


\begin{IEEEproof}[Proof of Proposition~\ref{prop:finite_sample_bnds}]
  Suppose $0\leq \alpha < 1.$ Then:
  \begin{align*}
    & \Pr\left\{L(\ROC, {\cal R}_{URC}(\hat{F},\alpha) ) \geq \delta\right\} \\
    & \stackrel{(a)}{=} \Pr\left\{L({\cal R}_{URC}(F,\alpha), {\cal R}_{URC}(\hat{F},\alpha) ) \geq \delta\right\} \\
    & \stackrel{(b)}{\leq} \Pr\left\{d_{KS}(F,\hat{F}) \geq (1-\alpha) \delta\right\} \\
    & \stackrel{(c)}{\leq}  2\exp\left( - 2 n(1-\alpha)^2\delta^2 \right) \\
  \end{align*}
  where (a) follows from ${\cal R}_{URC}(F,\alpha) = \ROC,$ (b) follows from Lemma~\ref{lemma:est_contraction}, and (c) follows from the DKW bound. This establishes \eqref{eq:UR_est_bnd} and the proof of \eqref{eq:LL_est_bnd} is similar.  The bound \eqref{eq:S_est_bnd} follows because it reduces to \eqref{eq:UR_est_bnd} if $0 \leq \alpha < 0.5$ and to \eqref{eq:LL_est_bnd} if $0.5 < \alpha \leq 1.$

  If $\alpha\in \{0,1\}$ then the fused and split estimators are the same so that in that case \eqref{eq:F_est_bnd} follows from \eqref{eq:S_est_bnd}.  It remains to prove \eqref{eq:F_est_bnd} assuming $0<\alpha < 1.$ Recall that the \levy\ metric is proportional to the $L^{\infty}$ metric for the functions rotated clockwise by $45^o.$ This fact and Lemma~\ref{lemma:est_contraction} imply:
  \begin{align*}
    L\left(\ROC, {\cal R}_{F}(\hat{F},\alpha)\right) & \leq \alpha L\left(\ROC, {\cal R}_{LLC}(\hat{F},\alpha)\right) + (1-\alpha) L\left(\ROC, {\cal R}_{URC}(\hat{F},\alpha)\right) \\
    & \leq \frac{\alpha}{\alpha}d_{KS}(F,\hat{F}) + \frac{1-\alpha}{1-\alpha}d_{KS}(F, \hat{F}) =2 d_{KS}(F,\hat{F})
  \end{align*}
  Thus, by the DKW inequality,
  \begin{align*}
    \Pr\left\{L(\ROC, {\cal R}_{F}(\hat{F},\alpha) ) \geq \delta\right\} \leq \Pr\left\{d_{KS}(F, \hat{F}) \geq \frac \delta 2\right\} \leq 2\exp(-n\delta^2/2),
  \end{align*}
  as was to be proved.
\end{IEEEproof}


\begin{thebibliography}{10}
\providecommand{\url}[1]{#1}
\csname url@samestyle\endcsname
\providecommand{\newblock}{\relax}
\providecommand{\bibinfo}[2]{#2}
\providecommand{\BIBentrySTDinterwordspacing}{\spaceskip=0pt\relax}
\providecommand{\BIBentryALTinterwordstretchfactor}{4}
\providecommand{\BIBentryALTinterwordspacing}{\spaceskip=\fontdimen2\font plus
\BIBentryALTinterwordstretchfactor\fontdimen3\font minus \fontdimen4\font\relax}
\providecommand{\BIBforeignlanguage}[2]{{%
\expandafter\ifx\csname l@#1\endcsname\relax
\typeout{** WARNING: IEEEtran.bst: No hyphenation pattern has been}%
\typeout{** loaded for the language `#1'. Using the pattern for}%
\typeout{** the default language instead.}%
\else
\language=\csname l@#1\endcsname
\fi
#2}}
\providecommand{\BIBdecl}{\relax}
\BIBdecl

\bibitem{Cox75}
D.~R. Cox, ``Partial likelihood,'' \emph{Biometrika}, vol.~62, no.~2, pp. 269--276, Aug. 1975.

\bibitem{KangHajek21}
\BIBentryALTinterwordspacing
X.~Kang and B.~Hajek, ``Lower bounds on information requirements for causal network inference,'' \emph{CoRR}, vol. abs/2102.00055, 2021. [Online]. Available: \url{http://arxiv.org/abs/2102.00055}
\BIBentrySTDinterwordspacing

\bibitem{Bradley97}
A.~P. Bradley, ``The use of the area under the {ROC} curve in the evaluation of machine learning algorithms,'' \emph{Pattern Recogn.}, vol.~30, no.~7, pp. 1145--1159, 1997.

\bibitem{MetzPan99}
C.~E. Metz and X.~Pan, ```{P}roper' binormal {ROC} curves: Theory and maximum-likelihood estimation,'' \emph{J. Math. Psychol.}, vol.~43, no.~1, pp. 1--33, Mar. 1999.

\bibitem{HsiehTurnbull96}
F.~Hsieh and B.~W. Turnbull, ``Nonparametric and semiparametric estimation of the receiver operating characteristic curve,'' \emph{Ann. Stat.}, vol.~24, no.~1, pp. 25--40, Feb. 1996.

\bibitem{Darlington73}
R.~Darlington, ``Comparing two groups by simple graphs,'' \emph{Psychological Bulletin}, vol.~79, no.~2, pp. 110--116, 1973.

\bibitem{Bamber75}
D.~Bamber, ``The area above the ordinal dominance graph and the area below the receiver operating characteristic graph,'' \emph{J. Math. Psychol.}, vol.~12, no.~4, pp. 387--415, 1975.

\bibitem{DeLongDeLongClarke-Pearson88}
E.~R. DeLong, D.~M. DeLong, and D.~L. Clarke-Pearson, ``Comparing the areas under two or more correlated receiver operating characteristic curves: A nonparametric approach,'' \emph{Biometrics}, vol.~44, no.~3, p. 837, Sep. 1988.

\bibitem{Aldrich97}
J.~Aldrich, ``{R.A. Fisher and the making of maximum likelihood 1912-1922},'' \emph{Statistical Science}, vol.~12, no.~3, pp. 162 -- 176, 1997.

\bibitem{Cramer46}
H.~Cram\'{e}r, \emph{Mathematical Methods of Statistics}.\hskip 1em plus 0.5em minus 0.4em\relax Princeton University Press, 1946, vol.~9.

\bibitem{Kang22}
\BIBentryALTinterwordspacing
X.~Kang, ``{ML} estimator of optimal {ROC} curve simulations,'' Feb. 2022. [Online]. Available: \url{https://github.com/Veggente/mleroc}
\BIBentrySTDinterwordspacing

\bibitem{GoelaRaginsky20}
N.~Goela and M.~Raginsky, ``Channel polarization through the lens of {B}lackwell measures,'' \emph{{IEEE} Trans. Inf. Theory}, vol.~66, no.~10, pp. 6222--6241, Oct. 2020.

\bibitem{Ben-TalNemirovski13}
A.~Ben-Tal and A.~Nemirovski, ``Optimization {III}: Convex analysis, nonlinear programming theory, nonlinear programming algorithms,'' 2013, \url{https://www2.isye.gatech.edu/~nemirovs/OPTIII_LectureNotes2018.pdf}.

\bibitem{WongHajek85}
E.~Wong and B.~Hajek, \emph{Stochastic Processes in Engineering Systems}.\hskip 1em plus 0.5em minus 0.4em\relax Springer New York, 1985.

\end{thebibliography}
\end{document}